\documentclass[a4paper,fleqn,usenatbib]{mnras}

\usepackage{newtxtext,newtxmath}

\usepackage{amsmath}
\usepackage{comment}
\usepackage{savesym}
\savesymbol{tablenum}
\usepackage[T1]{fontenc}
\usepackage{ae,aecompl}
\usepackage{subcaption}
\usepackage{color}
\newcommand{\sm}{\textsubscript{\(\odot\)}}
\newcommand{\brunt}{Brunt--V\"ais\"al\"a}
\usepackage{siunitx}
\restoresymbol{SIX}{tablenum}
\DeclareMathAlphabet{\mathitbf}{OML}{cmm}{b}{it}
\newcommand{\PD}[2]{\frac{\partial #1}{\partial #2}}
\renewcommand{\vec}{\mathitbf}

\newcommand{\msol}{\ensuremath{\mathrm{M}_\odot}}

\newcommand{\micromu}{\si{\micro}}

\usepackage{graphicx}	
\usepackage{amsmath}	
\usepackage{amssymb}	

\title[IGWs in Radiation Zones]{Two-Dimensional Simulations of Internal Gravity Waves in The Radiation Zones of Intermediate-Mass Stars}

\author[Ratnasingam et al.]{
R.~P. Ratnasingam,$^{1}$\thanks{r.p.ratnasingam2@newcastle.ac.uk}
P.~V.~F. Edelmann,$^{1,2}$
T.~M. Rogers$^{1,3}$
\\
$^{1}$School of Mathematics, Statistics and Physics, Newcastle University, Newcastle upon Tyne, NE1 7RU, UK\\
$^{2}$ X Computational Physics (XCP) Division and Center for Theoretical Astrophysics (CTA),  Los Alamos National Laboratory,  Los Alamos, NM 87545, USA\\
$^{3}$Planetary Science Institute, Tucson, AZ 85721, USA\\
}

\date{Accepted XXX. Received YYY; in original form ZZZ}

\pubyear{2018}

\begin{document}
\label{firstpage}
\pagerange{\pageref{firstpage}--\pageref{lastpage}}
\maketitle

\begin{abstract}
Intermediate-mass main sequence stars have large radiative envelopes overlying convective cores. This configuration allows internal gravity waves (IGWs) generated at the convective-radiative interface to propagate towards the stellar surface. The signatures of these waves can be observed in the photometric and spectroscopic data from stars. We have studied the propagation of these IGWs using two-dimensional fully-non-linear hydrodynamical simulations with realistic stellar reference states from the one-dimensional stellar evolution code, Modules for Stellar Astrophysics (MESA). When a single wave is forced, we observe wave self-interaction. When two waves are forced, we observe non-linear interaction (i.e. triadic interaction) between these waves forming waves at different wavelengths and frequencies. When a spectrum of waves similar to that found in numerical simulations is forced, we find that the surface IGW frequency slope is consistent with recent observations. This power law is similar to that predicted by linear theory for the wave propagation, with small deviations which can be an effect of nonlinearities. When the same generation spectrum is applied to 3 M$_{\odot}$ models at different stellar rotation and ages, the surface IGW spectrum slope is very similar to the generation spectrum slope.   

\end{abstract}

\begin{keywords}
stars:massive -- stars:interior -- waves
\end{keywords}



\section{Introduction}\label{sec:intro}
The propagation of internal gravity waves, or IGWs, within the confinement of the Earth's atmosphere is a well-researched topic \citep{2002AnRFM..34..559S,2003RvGeo..41.1003F,sutherland2010internal}. However, their properties in celestial objects and in particular, stars, are not clearly understood. Most importantly, the role of IGWs in the transport of angular momentum and energy within the radiation zones of stars has been the focus of significant research for a variety of stellar systems such as solar-like stars with convective envelopes and radiative cores \citep{2005MNRAS.364.1135R,2010MNRAS.404.1849B}, more massive stars with radiative envelopes and convective cores (\cite{2013ApJ...772...21R}, \cite{philipp3dpaper}) and evolved stars \citep{2014ApJ...796...17F}.

The generation of a broad spectrum of IGWs in stars has been mainly attributed to convective processes. One of the earliest works on this was by \cite{1981ApJ...245..286P}, which investigated the formation of IGWs through Reynolds stress arising from turbulent convection. The source terms for generation were originally derived in \cite{goldreich1977solar} for sound waves. Many works that followed such as \cite{1990ApJ...363..694G}, \cite{1999ApJ...520..859K} and \cite{2013MNRAS.430.2363L} extended this formulation. Another possible IGW generation model is the plume model. In a more recent publication, \cite{pincon2016a} theoretically investigated how convective plumes generate an IGW spectrum as a function of plume size and incursion time. \cite{philipp3dpaper} found that three-dimensional numerical simulations of a star with a convective core and a radiative envelope agreed with the theoretical spectrum in \cite{pincon2016a}.

IGWs in stars can either be inward-traveling waves or outward-travelling waves, depending on the location of the convection zones. At zero-age main sequence (ZAMS), stars up to 1.6 M$_{\odot}$ have radiative cores and convective envelopes while stars with masses above 1.6 M$_{\odot}$ tend to have radiative envelopes and convective cores. Higher mass stars usually develop intermediate or surface convection zones as they age, which also produce inward-traveling IGWs. In this work, however, we consider only stars with convective cores and radiative envelopes.

When considering only linear propagation of IGWs in stars, they are expected to undergo two main processes in the linear regime; amplitude increase due to density stratification and damping due to radiative diffusion. The amplitude also varies due to the spreading out of waves (geometric effect) and the \brunt{} frequency, but to a lesser degree. The collection of effects mentioned above affects the shape of the surface IGW spectrum. If the amplitude of a particular IGW exceeds a nonlinearity threshold, this wave can break, leading to a large amount of angular momentum transfer and mixing. \cite{1981ApJ...245..286P} combined the linearised versions of the Bousinesseq equations and anelastic equations with Wentzel-Kramers-Brillouin (WKB) theory to formulate a linear propagation equation for IGWs in stars, with radiative diffusion taken into account. \cite{1997A&A...322..320Z} used this formulation to investigate the angular momentum transport in the solar interior. More recently, \cite{rathish2019} showed that with the application of linear theory to IGWs, the fraction of waves that become non-linear varies depending on the stellar mass, age, metallicity and generation spectra for intermediate-mass to massive stars. 

When nonlinearity is considered, one of the possible interactions that can occur is triadic interaction. Generally, triad interactions refer to a two-wave interaction to generate a third wave. When this process occurs at a very high rate due to the interacting wave amplitudes being large, interaction between the third wave and the  initial waves produces even more waves rapidly, which can be considered as the wave breaking process. \cite{2010MNRAS.404.1849B} showed that tidal interactions between a solar-like star and a planet in an orbit can cause efficient wave breaking when a particular criterion is satisfied. \cite{2012ApJ...758L...6R} showed that the IGWs and their non-linear interactions of IGW can contribute to the reversal of stellar surface rotation.

The main aim of this work is to investigate the properties of IGWs at the surface of stars with radiative envelopes and convective cores. It will serve as an extension to \cite{rathish2019}, which considered linear IGWs only. We achieve this aim by running two-dimensional (2D) annulus simulations of the radiative zones with different IGW generation spectra. The background reference states are obtained from the one-dimensional (1D) stellar evolution code MESA \citep{2011ApJS..192....3P}. This work uses realistic background stellar thermal diffusivities, whilst previous works on IGW simulation \citep{2005MNRAS.364.1135R,2013ApJ...772...21R,philipp3dpaper} use much higher thermal diffusivities to maintain numerical stability. Furthermore, we were able to use the full extent of the radiation zone, excluding sub-surface convection zones, which has not been done in previous work due to the large density variation from the core to the surface. 

This paper is structured into 5 sections. Section~\ref{sec:background} is on background theory which connects the non-linear simulations done for this paper to the work done on linear theory. This section ends with a subsection on the different generation spectra of IGWs tested in this work. Section~\ref{sec:resilve_IGW} explains the work done to resolve the IGWs in our simulations. Section~\ref{sec:results} describes our results for monochromatic wave simulations, two-wave simulations and multiple wave simulations. The final section of the paper summarises and concludes this work.

\section{Background Theory and Methods}\label{sec:background}
\subsection{Nonlinear Simulations}
Our non-linear simulations were run using a pseudo-spectral code written for solving the Navier-Stokes equations in the anelastic approximation (Eq.~\eqref{eq:anelastic-rho} -- Eq.~\eqref{eq:anelastic-T}), within the geometry of an equatorial slice of the chosen stellar model \citep{2005MNRAS.364.1135R}. The equations are solved using a finite-difference scheme in the radial direction and the Fourier spectral method in the azimuthal direction. The Adams-Bashforth timestepping method is applied to the non-linear terms while the Crank-Nicolson method is applied to the linear terms. The equations are
\begin{align}
\label{eq:anelastic-rho}
\nabla \cdot \overline{\rho} \vec{v} = 0,
\end{align}
\begin{align}
\label{eq:anelastic-v}
\PD{\vec{v}}{t}&= - (\vec{v} \cdot \nabla) \vec{v}
- \nabla \left(\frac{P}{\overline{\rho}}\right) - C \overline{g} \vec{\hat{r}} + 2(\vec{v} \times \vec{\hat{z}} \Omega) \\
\nonumber
& + \overline{\nu} \left( \nabla^2 \vec{v} + \frac{1}{3} \nabla (\nabla \cdot \vec{v}) \right),\\
\label{eq:anelastic-T}
\PD{T}{t}&=  - (\vec{v} \cdot \nabla) T + (\gamma - 1) T h_\rho v_r\\
\nonumber
&- v_r \left( \PD{\overline{T}}{r} - (\gamma - 1) \overline{T} h_\rho \right)\\
\nonumber
& + \frac{1}{c_v\overline{\rho}} \nabla \cdot (c_p \overline{\kappa}\overline{\rho}\nabla T) + \frac{1}{c_v\overline{\rho}} \nabla \cdot (c_p \overline{\kappa}\overline{\rho}\nabla \overline{T}).
\end{align}
where $\rho$, $T$ and $P$ are the perturbation density, temperature and pressure respectively whilst the same quantities with an over-line represent reference state quantities which vary only with radius.  velocity, $v_r$ and the tangential velocity, $v_{\theta}$, together form the velocity vector, $\vec{v}$. The negative inverse density scale height is represented by $h_\rho$. The rotational angular velocity and viscosity are represented by $\Omega$ and $\overline{\nu}$. The adiabatic constant, $\gamma$ is set to be 5/3. The specific heat capacity at constant volume, specific heat capacity at constant pressure and thermal diffusivity are represented by $c_v$, $c_p$ and $\overline{\kappa}$ respectively. The reference state gravity is represented by $\overline{g}$. The co-density, $C$, is 
\begin{equation}
\label{eq:codensity}
C=-\frac{1}{\overline{T}}\left(T  + \frac{1}{\overline{g} \overline{\rho}}\PD{\overline{T}}{r}P \right).
\end{equation}
The temperature perturbation, $T$, was set to zero at both the bottom and top boundaries of the simulation domain. The radial velocities, $v_r$, were set to zero at the top boundary and forced at the bottom boundary to mimic wave generation. For the horizontal velocities, $v_{\theta}$ at the top boundary, a stress-free boundary condition was imposed. At the bottom boundary, the divergence-free mass-flux condition (see Eq.~\eqref{eq:anelastic-rho}) ensures that $v_{\theta}$ is forced because in our simulations, we set $v_r$ through the streamfunction, $\psi$ (see Eq.~\eqref{eq:vgen} in Section~\ref{sec:gen_spec}).  

To simplify computations, we introduce a streamfunction, $\psi$, which is defined through its relation with the radial and horizontal velocities:
\begin{equation} \label{eq:vr_vphi}
v_r = \frac{1}{\overline{\rho}r}\frac{\partial \psi}{\partial \theta} ; \;
v_{\theta} = -\frac{1}{\overline{\rho}}\frac{\partial \psi}{\partial r}.
\end{equation} 
This allows us to write Eq.~\eqref{eq:anelastic-v} in terms of vorticity, defined as $\nabla \times \vec{v}$, which reduces the number of equations needed to be solved and ensures that the divergence-free mass flux criterion is fulfilled by construction.

We use MESA to generate a models of a 3 $\mathrm{M}_{\odot}$ star at various ages, which are used as the reference state for our simulations. Stellar metallicity, $Z$, was set to be equal to the solar value of $Z$ = 0.02. The mixing length parameter was set as 1.8 and the convective overshoot profile was set to exponential. The \brunt{} frequency, $N$, was calculated as in \cite{2013ApJ...772...21R}: 
\begin{equation}
N^2 = \frac{\overline{g}}{\overline{T}}\left(\frac{d\overline{T}}{d\overline{r}}-\left(\gamma-1\right)\overline{T}h_{\rho}\right).
\end{equation}
The sign of the \brunt{} frequency was used to determine the extent of the radiation zone defined by 
\begin{equation}
N(r)^2 > 0 .
\end{equation}
This means that for all the models that we have used, the simulation domain starts at the top of the convective core and ends at the base of the surface convection zone. For a 3 M$_{\odot}$ star at ZAMS, this covers the range from 16\% to 99\% of the total stellar radius. In our simulations, this whole region is divided into 1500 grids in the radial direction and 1024 grids in the horizontal direction.  

\subsection{Linear Theory}\label{sec:linear_theory}
Application of linear theory to investigate IGWs has been done extensively in the field of oceanography \citep{sutherland2010internal}. The general idea is to linearise hydrodynamical equations and introduce a wave-like ansatz to solve the linearised equations. In the context of stellar parameters, one of the earliest works was done by \cite{1981ApJ...245..286P}, which focused on solar-like stars. In our work, we follow similar steps, starting with the linearised hydrodynamical equations in the anelastic approximation:

\begin{align}
\label{eq:anelastic-linear-rho}
\nabla \cdot \overline{\rho} \vec{v} = 0,
\end{align}
\begin{align}
\label{eq:anelastic-linear-v}
\PD{\vec{v}}{t}&= 
- \nabla \left(\frac{P}{\overline{\rho}}\right) - C \overline{g} \vec{\hat{r}} \\
\label{eq:anelastic-linear-T}
\PD{T}{t}&=  
- v_r \left( \PD{\overline{T}}{r} - (\gamma - 1) \overline{T} h_\rho \right) 
\end{align}
We ignore the pressure term in eq.~\eqref{eq:codensity}, which leads to a set of equations that conserve energy \citep{Brown2012}. We also do not consider rotational, thermal diffusion or viscous effects. In cylindrical coordinates, this two-dimensional analysis allows us to set the z-derivatives to zero. The three equations shown above can then be reduced to one second-order differential equation with $v_r (r)$ as the evolving term:

\begin{align}\label{eq:linear_vr}
\nonumber
0 &= \frac{\partial^2 \alpha}{\partial r^2} + \left( \frac{N^2}{\omega^2} - 1 \right) \frac{m^2}{r^2} \alpha \\ 
\nonumber
&+ \left[ -\overline{\rho}^{-1/2}\frac{\partial^2 \left(\overline{\rho}^{1/2}\right)}{\partial r^2} + \frac{\partial h_{\rho}}{\partial r} \right]\alpha\\
&+ \frac{1}{4r^2}  \alpha
\end{align}
where $\alpha = v_r \overline{\rho}^{1/2} r^{3/2}$. We have used a wave ansatz of the form $v_r (r,\theta,z) \propto v_r(r) e^{im\theta}e^{-i\omega t}$, where $m$ is the horizontal wavenumber\footnote{We use the term, wavenumber to represent $m$ and $k_h$, which is $m/r$, interchangeably in this paper and in all cases, we make it clear which one we are referring to.}, $\omega$ is the angular frequency and $\theta$ is the angular coordinate in the equatorial plane. Generally, in the radiation zone, there will be regions where the oscillatory term (OT), $\left(N^2/\omega^2 - 1 \right) m^2/r^2$, dominates and regions where the density term (DT), $\left[ -\overline{\rho}^{-1/2}\partial^2 \left(\overline{\rho}^{1/2}\right)/\partial r^2 + \partial h_{\rho}/\partial r \right]$, dominates. When the ratio of OT to DT is less than 1, an IGW loses its wavelike behaviour and the approximate radius where this ratio is exactly equal to 1 is called the turning point. The importance of this will be discussed in Section~\ref{sec:results}.

To investigate the linear behaviour of IGWs in the radiation zone, we neglect the DT\footnote{With the assumption that IGW wavenumber varies much more rapidly than the background density, which is equal to a locally Boussinesq approximation.} and the last term\footnote{This term is always smaller than $\left(N^2/\omega^2 - 1 \right) m^2/r^2$ as $\omega$ is always smaller than $N$, and $m$ is always bigger than 1.} in Eq~\eqref{eq:linear_vr}. Applying the WKB approximation to the remaining form of the equation allows us to determine the dependence of radial IGW amplitude as follows:
\begin{equation}
v_r \propto \rho^{-1/2} \; r^{-1} \; (N^2 - \omega^2)^{-1/4}.
\end{equation} 
Applying the same procedure to the anelastic equations in three dimensions, using spherical harmonics, gives a similar dependence of $v_r$ on $\rho$ but slightly different dependence on $r$, as follows: 
\begin{equation}
v_r \propto \rho^{-1/2} \; r^{-3/2} \; (N^2 - \omega^2)^{-1/4}.
\end{equation} 
Comparing the results in 2D and 3D, we obtain a simple conversion factor of $r^{1/2}$. Therefore, in 2D, we expect waves to have higher surface amplitudes due to this geometric effect.  

\subsection{Generation Spectra}\label{sec:gen_spec}
In order to mimic the generation of IGWs by convection without the added computational cost, we investigate three different prescriptions for the spectrum of waves forced at the innermost boundary of the radiation zone. The waves are forced as perturbations to the radial velocity, which depend on the frequency and wavenumber in the following manner:
\begin{equation}\label{eq:spectra_proportionality}
v_{r,0}(\omega,\ell) \propto \omega^a  m^b
\end{equation}
where the values of $a$ and $b$ for different prescriptions are shown in Table~\ref{table:mandn} \citep{rathish2019}.For spectrum $\mathrm{R}_{\mathrm{break}}$, we have given two values for "a" as this spectrum is one with a broken power law with a transition at 30 \si{\micro}Hz. 

\begin{table}
	\centering		
	\begin{tabular}{ccc}
		\hline \hline
		Spectra & a & b \\ \hline \hline 
		\cite{1999ApJ...520..859K}[K] & -2.17 & 1 \\
		\cite{2013MNRAS.430.2363L}[LD] & -4.25 & 2.5 \\
		\cite{2013ApJ...772...21R}[R$_{\text{break}}$] & -0.6/-2.4 & -0.9 \\ \hline
		\end{tabular}
		\caption{The table shows how $a$ and $b$ in Eq.~\eqref{eq:spectra_proportionality},  are defined for different works \citep{rathish2019}. The flat spectrum represents one with no dependence on $\omega$ or $\ell$. The letters shown in square brackets will be used to represent its respective spectrum. For spectrum $\mathrm{R}_{\mathrm{break}}$, we have given two values for "a" as this spectrum is one with a broken power law with a transition at 30 \si{\micro}Hz. }
		\label{table:mandn}
\end{table}

The forcing of these spectra in our simulations was done using cosine functions, as shown below:
\begin{equation} \label{eq:vgen}
	\psi = A_{gen} r_c \rho_c \sum_{m=1}^{N_m} m^{b-1} \cos(2\pi mx) \sum_{i=1}^{N_f} \left(\frac{\omega_i}{\omega_c}\right)^a \cos(\omega_i t),
\end{equation}    
where $\psi$ is the same as in Eq.~\eqref{eq:vr_vphi}. The terms, $N_m$ and $N_f$ are the number of discrete wavenumbers and frequencies (constrained by the resolution of our simulations). The convective turnover frequency, $\omega_c$ is defined as $u_c/r_c$, where $u_c$ is the convective velocity provided by the mixing length theory (MLT) formulation in MESA and $r_c$ is the extent of the convection zone. For a 3\msol{} star at ZAMS, this value is  approximately 0.0485 \si{\micro}Hz and varies insignificantly between the models we have used. The stellar density at the bottom of the radiation zone is defined as $\rho_c$ respectively. To keep the energy input for the different generation spectrum constant, we introduce the constant, $A_{gen}$. This value is calculated from matching the volume-averaged mixing length velocity in the convection zone, provided by MESA, with the integrated $v_{r,gen}$ over all the tested wavenumbers and frequencies. 

We work primarily with Spectrum R$_{\mathrm{break}}$ in this paper for reasons explained in Section~\ref{sec:results}. One of the concerns with using this type of forcing is whether it matches the expected convective flux, given that the convective turnover frequency is 0.0485 \micromu{}Hz. In \cite{2013ApJ...772...21R}, the wave flux was found to be roughly $10^{-2}$ of the convective flux, $ F_{\mathrm{conv}} = \rho_c u_c^3$. Using the work done in \cite{1997A&A...322..320Z}, \cite{kiraga2003direct} and \cite{2014A&A...565A..42A}, the total integrated wave flux is calculated as follows:
\begin{align}\label{eq:Fw}
F_w &= \int_{\omega}^N\int_{m} \rho \frac{v_r^2(k_h,\omega)}{k_h \omega} V_{g}(k_h,\omega) \; \mathrm{d}m \; \mathrm{d}\omega\\
&= \int_{\omega}^N\int_{m} \frac{\rho r}{N^2} \frac{\omega v_r^2\sqrt{N^2 - \omega^2}}{m^2} \; \mathrm{d}m \; \mathrm{d}\omega, 
\end{align}
where $V_g$ is the group velocity. To account for how wave forcing is a smooth process which occurs over the convective-radiative interface, we calculated the integrated wave flux close to the bottom boundary of the simulation domain (and not at the exact boundary), for all of our models, and present them in Table~\ref{tab:flux}. The integrated wave flux was found to be consistent with the results seen in \cite{2013ApJ...772...21R}. For the monochromatic wave forcing, these are test cases, which means we do no expect realistic stellar convection of a 3\msol{} star to produce waves with such amplitudes/fluxes.

\begin{table}
	\centering
	\begin{tabular}{cc}
		\hline \hline
		Model   & Integrated Flux (Units of $F_{\mathrm{conv}}$) \\
		\hline \hline
		ZAMS (no rotation) & 0.025\\
		ZAMS ($\Omega = 0.796$\micromu{}Hz) & 0.016\\
		ZAMS ($\Omega = 12.73$\micromu{}Hz) & 0.021\\
		midMS & 0.049 \\
		TAMS & 0.094\\
		\hline
	\end{tabular}
	\caption{The integrated flux for all the models in this work calculated using Eq.~\eqref{eq:Fw}. Both $\Omega$ in the table refer to the rotation rate of the models, whilst midMS and TAMS refer to middle-of-the-Main-Sequence and Terminal-Age-Main-Sequence models respectively.}
	\label{tab:flux}
\end{table}

\section{Resolving Internal Gravity Waves}\label{sec:resilve_IGW}
We want to limit our conclusions and investigation to waves which can be both resolved numerically and do not dissipate completely due to thermal diffusion and viscosity. First, we discuss the resolution limit. Low-frequency IGWs have short radial wavelengths, which can be difficult to resolve numerically. We estimate the resolving power of our simulations using the dispersion relation of IGWs, which is
\begin{equation} \label{eq:kr}
k_r = \frac{m}{r}\sqrt{\left( \frac{N^2}{\omega^2} -1\right)}, 
\end{equation}
where $k_r$ is the radial wavenumber of an IGW. The radial wavelength, $\lambda_r$, of an IGW is then $2\pi/k_r$. To find the grid spacing required to resolve an IGW in the vertical direction, we calculate the ratio of radial wavelength to grid spacing, $dr$, for a given horizontal wavenumber, $m$, and angular frequency, $\omega$,  using
\begin{equation} \label{eq:ratio_lambda_r_over_dr}
\frac{\lambda_r}{dr} = \frac{1}{dr}\frac{2\pi r}{m}\sqrt{\frac{\omega^2}{N^2 - \omega^2}}.
\end{equation}

\begin{figure}
	\centering
	\includegraphics[trim={0.0cm 0.0cm 0 0.0cm},clip,width=\columnwidth]{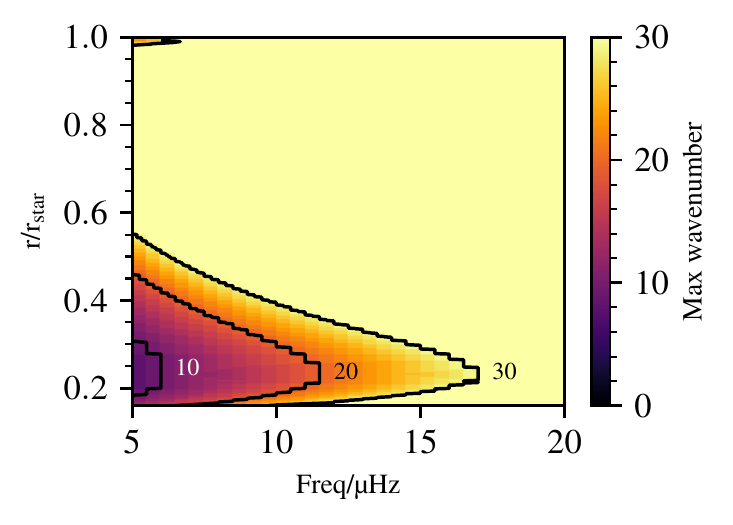}
	\caption{The smallest radial wavelengths that can be resolved with four grid points as a function of radius and wave frequency in \si{\micro}Hz. The contour lines represent regions where the maximum wavenumber is equal to the value indicated next to the lines.\label{fig:max_wavenumber_pcolor}}
	\centering
\end{figure}

Figure~\ref{fig:max_wavenumber_pcolor} shows the maximum IGW wavenumber that can be resolved for a frequency range of 5 \si{\micro}Hz to 100 \si{\micro}Hz inside the propagation cavity, using a minimum resolution requirement of four grid cells per vertical wavelength, $\lambda_r/dr = 4$. The smaller resolving power at lower frequencies is due to lower frequency waves having smaller vertical wavelengths. The increase in resolving power with increasing radius is due to increasing vertical wavelengths, caused by decreasing \brunt{} frequency (see Section~\ref{sec:results}). Very close to the surface, the \brunt{} frequency increases very rapidly before dropping to zero, causing the resolving power to decrease. From the plot, one can see that we cannot reliably resolve lower frequency and larger wavenumber waves, especially at small radii. Applying a similar procedure in the horizontal direction shows that even the smallest wavelengths used in this paper are resolved by more than 50 grid points. 

Another limit affecting IGWs is given by thermal diffusion and viscosity. The latter in particular had to be increased beyond its stellar values for numerical reasons. We expect IGWs with high wavenumbers and low frequencies to be affected primarily and we use a dimensional analysis argument to roughly estimate the largest IGW wavenumber before an IGW is completely dissipated by viscous effects/thermal diffusion. This means to match the dimensions of viscosity/thermal diffusivity, we use the wavelength and frequency. We chose the total wavelength corresponding to the total wavenumber, $\left(k_r^2 + k_h^2\right)^{0.5}$ and the angular wave frequency, $\omega$ respectively and under the assumption that $k_r >> k_h$, we obtain
\begin{equation}
m_{max} =  \left(\frac{2\pi}{\max(\nu,\kappa)}\frac{r^2\omega^3}{N^2 - \omega^2}\right)^{1/2},
\end{equation}
where $m_{max}$ is the maximum wavenumber before waves are completely dissipated by thermal diffusion or viscosity and the $\max$ function selects the larger of the two arguments. Along the radial path of a 10 \micromu Hz wave, the minimum value of $m_{max}$ is found to be 6. For all frequencies above 10 \micromu Hz, $m_{max}$ was found to be higher and thus, taking into account the numerical resolution limit and the thermal diffusion/viscosity limit, we chose an optimal frequency range of 10 \si{\micro}Hz to 500 \si{\micro}Hz and a wavenumber range of 1 to 20 for all analysis in the following sections. Note that this range is dependent on the type of the radial and angular discretisation (i.e. finite-difference and spectral) we have used. 

\section{Results} \label{sec:results}
\subsection{Monochromatic Wave Analysis}
\subsubsection{Single wave validation}
As IGWs propagate through the stably-stratified layer of a star, they experience numerous effects: they are affected by the density stratification, dissipated by thermal diffusion and they interact with the varying \brunt{} frequency. They may also experience non-linear wave-wave interactions and if their amplitudes are sufficiently large, wave breaking may occur.
 
\begin{figure*}
	\centering
	\includegraphics[trim={0.0cm 0.0cm 0 0.0cm},clip,width=0.9\textwidth]{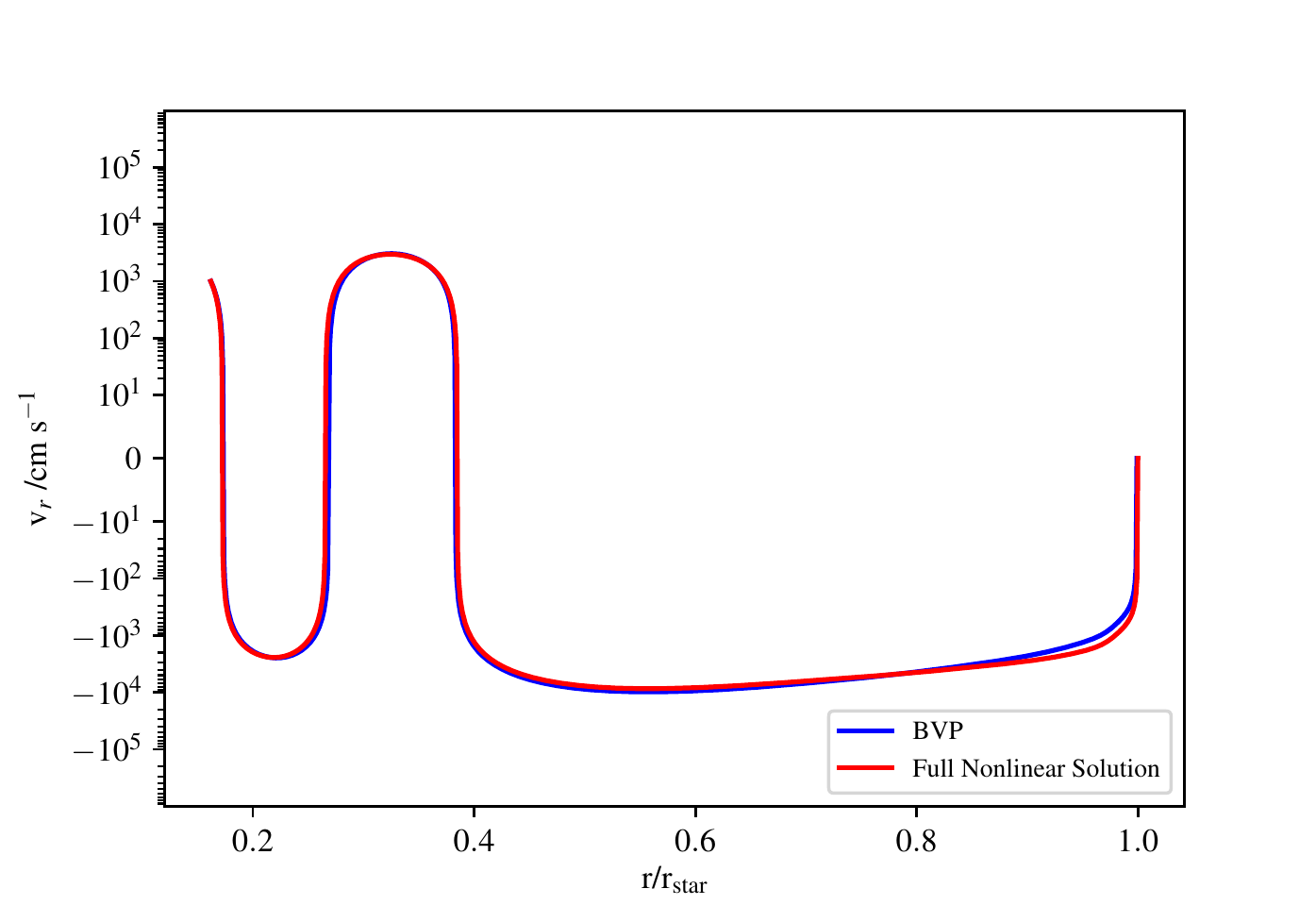}
	\caption{Radial velocities against the radius of the star in units of the total stellar radius. The red line represents solutions from the fully non-linear hydrodynamical simulation. The blue line represents solution from solving Eq.~\eqref{eq:anelastic-linear-v} as a boundary value problem (BVP). The y-axis of the plot is a symmetric log plot which uses linear scaling between -10 and 10, and log scaling outside this range.  \label{fig:single_wave_validity_check}}
	\centering

\end{figure*}
To start off our investigation on IGWs in the radiation zone, we forced single monochromatic IGWs at the bottom boundary of the domain. We set the background viscosity to a constant value $1 \times 10^{12}$ cm$^2$ s$^{-1}$. The simulations were run for a time of 720 wave cycles for a forced wave with an amplitude of 1000 cm s$^{-1}$, frequency of 120 \si{\micro}Hz and wavenumber of $m = 3$. Note that the magnitude of the amplitude was chosen based on the range for bulk velocities (100 cm s$^{-1}$ to 10000 cm s$^{-1}$) in the convection zone predicted by MLT for this stellar model. 

Figure~\ref{fig:single_wave_validity_check} shows the radial velocity profile of a 120 \micromu Hz, $m=3$ wave as a function of radius. The red line represents results from the fully non-linear simulation while the blue line represents the solution from Eq.~\eqref{eq:linear_vr} solved using a tridiagonal matrix solver with the top and bottom boundary values set to match the boundary values of the hydrodynamical simulation. Although the simulation is fully nonlinear, the good matching between the red line and the blue line here can be attributed to three factors. First, the forced wave experiences very little radiative damping due to its high frequency \citep{rathish2019}. Second, varying the explicit viscosity in the domain produced no observable changes to the waveform, meaning it is not in the regime where viscous damping is relevant. Finally, the forced wave has a low nonlinearity parameter, 
\begin{equation}\label{eq:epsilon}
	\epsilon = k_r \frac{v_r}{\omega},
\end{equation} 
as defined in \cite{rathish2019}. We find that $\epsilon$ does not exceed 0.00127 anywhere in the domain. Thus, we do not expect the forced wave to experience strong non-linear effects ($\epsilon = 1$). However, these waves might still undergo weak non-linear interactions, as seen in later sections.  

\begin{figure*}
	\centering
	\includegraphics[trim={0.0cm 0.0cm 0 0.0cm},clip,width=0.8\textwidth]{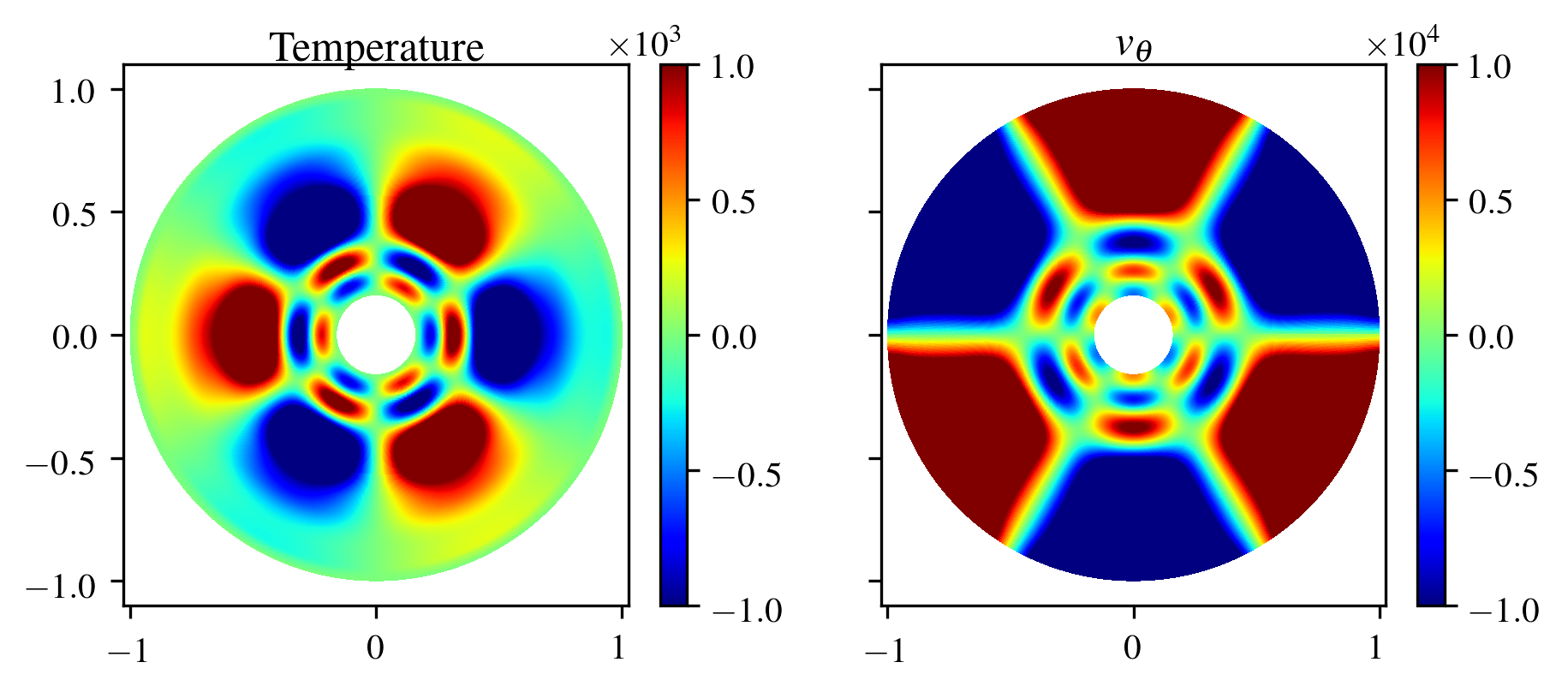}
	\caption{Colour plots of different parameters at a $t = 60$ wave cycles for a wave forced at $f$ = 120 \si{\micro}Hz and $m = 3$; temperature (in Kelvin) and horizontal velocity (in cm s$^{-1}$). The x and y -axes represent the cartesian x and y coordinates in units of total stellar radius. The temperature and horizontal velocities can be seen to be out of phase by 90 degrees.   \label{fig:snapshot112}}
	\centering

\end{figure*}

\subsubsection{Nonlinear simulations of monochromatic waves}\label{sec:nonlin_mono_sim}
Figure~\ref{fig:snapshot112} shows a time snapshot of temperature perturbation and horizontal velocity profiles for the same wave discussed in the previous section. We can see how forcing a single wave at the bottom boundary leads to the evolution of perturbation quantities forming wave-like structures. The temperature perturbations and the horizontal velocities are out of phase by $\pi / 2$ rad, as expected. 

\begin{figure}
	\centering
	\includegraphics[trim={0.0cm 0.0cm 0 0.0cm},clip,width=\columnwidth]{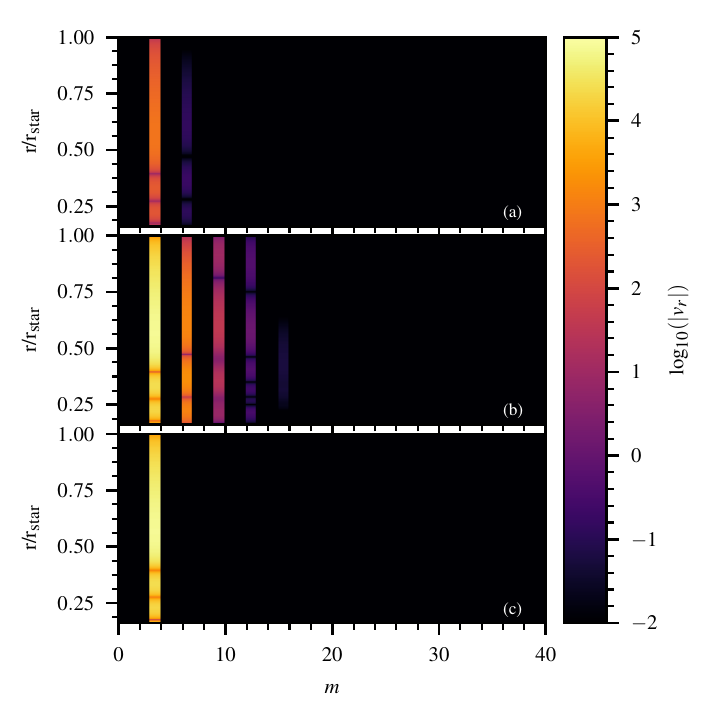}
	\caption{Spatial Fourier transform of the radial velocities at $t = 60$ wave cycles at different radii for a wave forced at 120 \si{\micro}Hz and $m = 3$. The top panel (a) represents simulations results with a wave forced at an amplitude of 100 $\mathrm{cm\; s^{-1}}$ whilst the middle panel (b) represents simulation results with a wave forced at an amplitude of 1000 $\mathrm{cm\; s^{-1}}$. The bottom panel (c) represents simulation results with no non-linear terms forced at an amplitude of 1000 $\mathrm{cm\; s^{-1}}$. \label{fig:hrfft_single_wave}}
	\centering 

\end{figure}

In a non-linear system, we expect forcing single waves at the bottom boundary to lead to  self-interaction. Thus, to investigate this, we look at how energy is transfered between waves of different wavenumbers. In Fig.~\ref{fig:hrfft_single_wave}, we show wave amplitudes as functions of stellar radius and wavenumbers for a forced wave at 120 \si{\micro}Hz and $m = 3$. The forced wave has an amplitude of 100 cm s$^{-1}$ in panel (a), 1000 cm s$^{-1}$ in panel (b) and 1000 cm s$^{-1}$ but with no non-linear terms in panel (c). In Fig.~\ref{fig:hrfft_single_wave}\textcolor{blue}{(a)}, we observe a wave at $m = 6$, which is an indication of energy transfer from larger wavelengths to smaller wavelengths. Despite having a very low nonlinearity parameter ($\epsilon \leq 0.000126$), there is still energy transfer occurring. Increasing the amplitude to 1000 cm s$^{-1}$ ($\epsilon \leq 0.00126$) causes more waves to be generated at larger $m$, as shown in Fig.~\ref{fig:hrfft_single_wave}\textcolor{blue}{(b)}, whilst switching off the non-linear terms completely removes waves at higher wavenumbers, as shown in Fig.~\ref{fig:hrfft_single_wave}\textcolor{blue}{(c)}. As expected, this shows that the non-linear wave self-interaction transfers energy from larger scales to smaller scales and the energy transfer rate is dependent on the forced wave amplitude.    

To quantify the rate of energy transfer as a function of wave amplitude, we compare the amplitudes of the higher harmonics for the two different forcing amplitudes. Table~\ref{table:amplitude_comparison} shows the ratio of harmonic amplitudes from the simulations with the stated forcing amplitude (100 cm s$^{-1}$, 1000 cm s$^{-1}$ and 10000 cm s$^{-1}$) to the harmonic amplitudes from the simulation with the lowest forcing amplitude (100 cm s$^{-1}$). At the zeroth harmonic ($m=3$) order, the ratio of harmonic amplitudes increases by the same order of magnitude as the increase in the forcing amplitude, as expected. At the first harmonic order ($m=6$), the ratio of amplitudes is 100 and at the second harmonic order ($m=9$), it is 1000. Running another simulation with a forcing of 10000 cm s$^{-1}$, we find the ratios of harmonic amplitudes amplitudes to be $10^4$ at $m = 6$ and $10^6$ at $m = 9$. This indicates that the energy transfer to higher wavenumbers is larger for higher forced wave amplitudes and it is proportional to $A_r^{2(h+1)}$, where $A_r$ is the ratio of amplitudes between the forced waves and $h$ is the harmonic order. The factor of 2 comes from the dimensional argument that wave energy is proportional to the square of velocity.
\begin{table}
	\centering		
	\begin{tabular}{c}
		\hline \hline
		 \hspace{3.5cm}Ratio of harmonic amplitudes \\ 
		 \begin{tabular}{cccc}
			Forcing Amplitude/cm s$^{-1}$ & $m=3$ & $m=6$ & $m=9$  \\ \hline \hline 
			100 & 1 & 1 & 1\\ 
			1000 & 10 & $10^2$ & $10^3$ \\
			10000 & 100 & $10^4$ & $10^6$ \\ \hline
		\end{tabular}
	\end{tabular}
	\caption{The table shows the approximate ratio of harmonic amplitudes between stated forcing amplitudes (left column) and the lowest forcing amplitude, which is 100 cm s$^{-1}$. These ratios are approximate averages over the whole simulation regime. The forced wavenumber is $m = 3$, whilst the first harmonic is $m=6$ and the second harmonic is $m=9$. }
	\label{table:amplitude_comparison}
\end{table}

\begin{figure}
	\centering
	\includegraphics[trim={0.0cm 0.0cm 0 0.0cm},clip,width=\columnwidth]{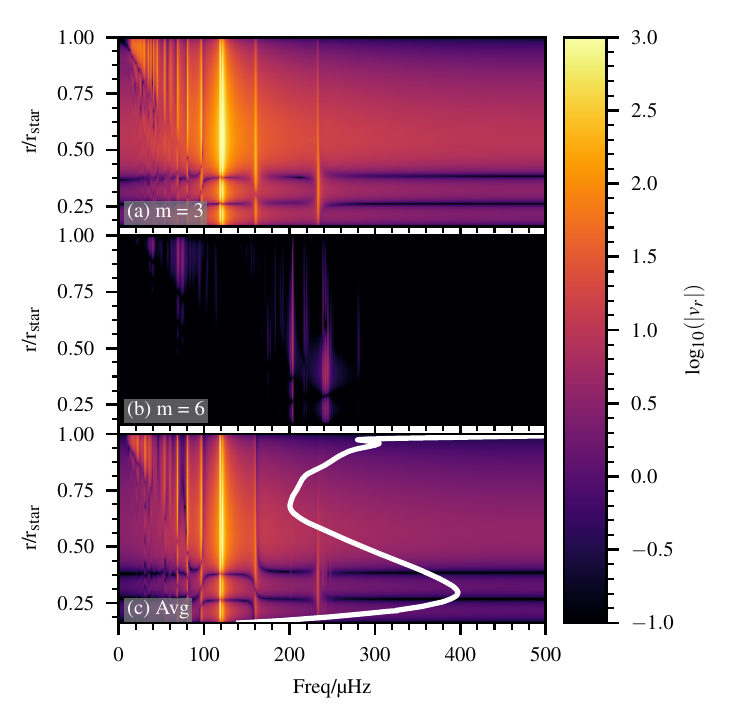}
	\caption{Radial velocities as functions of stellar radius and frequency for a wave forced at 120 \si{\micro}Hz and $m = 3$. Panel (a) shows the radial velocities at a wavenumber $m = 3$, panel (b) is for $m=6$ and panel (c) shows the wavenumber-averaged radial velocities. The white line in panel (c) indicates the \brunt{} frequency profile for this stellar model. We observe waves at frequencies lower than the forced wave frequency, which is likely related to the interaction between these waves and the varying \brunt{} frequency profile.\label{fig:trfft_single_wave}}
	\centering

\end{figure}

Moving to energy transfer between waves in frequency space, Fig.~\ref{fig:trfft_single_wave} shows vertical velocity amplitudes at different radii and frequencies for the $m = 3$ (Fig.~\ref{fig:trfft_single_wave}\textcolor{blue}{(a)}) and $m=6$ (Fig.~\ref{fig:trfft_single_wave}\textcolor{blue}{(b)}) as well as amplitudes averaged over all wavelengths (Fig.~\ref{fig:trfft_single_wave}\textcolor{blue}{(c)}), for the simulation with a 120 \si{\micro}Hz, $m=3$ wave, forced with an amplitude of 1000 cm s$^{-1}$. The \brunt{} frequency profile has been plotted (white line) in the bottom panel. We find two main results from this analysis. First, forcing a wave close to a cavity mode frequency\footnote{By cavity mode frequency, we mean the eigenfrequencies of the system. These frequencies result from the eigenmode solution of the linear Navier-Stokes equations. These are stationary waves which are also referred to as g-modes in asteroseismology.} produces more efficient energy transfer to different cavity mode frequencies. For example, from Fig.~\ref{fig:trfft_single_wave}\textcolor{blue}{(a)}, we can see that at approximately 235 \si{\micro}Hz, a fundamental mode (stationary wave with zero nodes) can be observed and at 160 \si{\micro}Hz, the first harmonic cavity mode (stationary wave with 1 node). Second, weakly non-linear interactions between waves allow energy transfer to waves at frequencies that are integer multiples of the forced wave frequency. In Fig.~\ref{fig:trfft_single_wave}\textcolor{blue}{(b)}, we observe a wave at 240 \si{\micro}Hz, which is indicative of triadic interactions. The peak with two nodes seen at 200 \si{\micro}Hz is likely to be the result of a daughter wave interaction with a cavity mode. 

\begin{figure}
	\centering
	\includegraphics[trim={0.0cm 0.0cm 0 0.0cm},clip,width=\columnwidth]{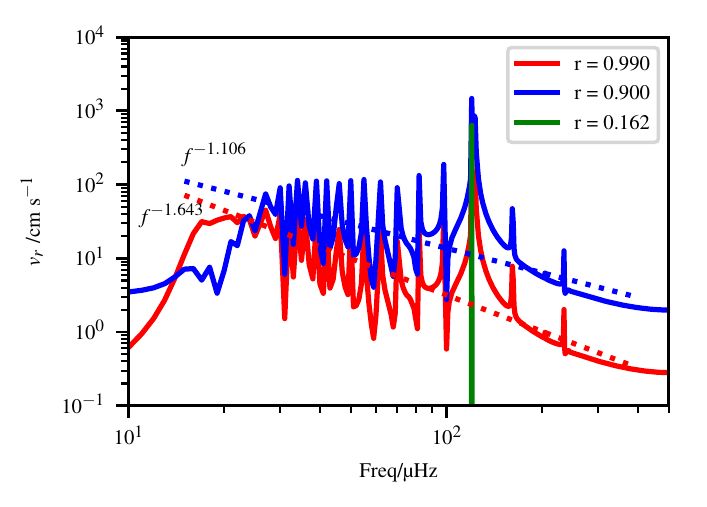}
	\caption{ Temporal Fourier transforms of the radial velocities at three different radii for a wave forced at 120 \si{\micro}Hz and $m = 3$. The radii given in the legend are in units of the total stellar radius. The dashed lines represent straight line fits to $\log_{10} (v_r)$ and $\log_{10}(\mathrm{freq})$.\label{fig:single_line_plot_surface}}
	\centering

\end{figure}

Figure~\ref{fig:single_line_plot_surface} shows line plots of the radial velocity spectrum at the top of the radiation zone (r/r$_{\mathrm{star}}=0.99$ and r/r$_{\mathrm{star}}=0.90$)\footnote{We will refer to r/r$_{\mathrm{star}}$ as r from here on unless stated otherwise.}. Past numerical simulations of stellar interiors that included the convective core and resolved IGWs have been limited to r = 0.9 for numerical stability, which is why we include the spectrum this radius for comparison purposes. The spike with the green line indicates the forcing frequency. The propagation of this wave to the surface causes similar peaks at r = 0.90 (blue line) and r = 0.99 (red line), as expected. The slope in logarithmic frequency space was found to be -1.106 at 90\% the total radius and -1.643 at 99\% the total radius.  At lower frequencies, thermal diffusion dominates, leading to waves being damped over a short distance. Thus, the presence of wave structures at low frequencies show that the transfer of energy from high frequency waves to low frequency waves occurs locally, as seen in Fig.~\ref{fig:trfft_single_wave}\textcolor{blue}{(a)} and  Fig.~\ref{fig:single_line_plot_surface}.

\begin{figure}
	\centering
	\includegraphics[trim={0.0cm 0.0cm 0 0.0cm},clip,width=\columnwidth]{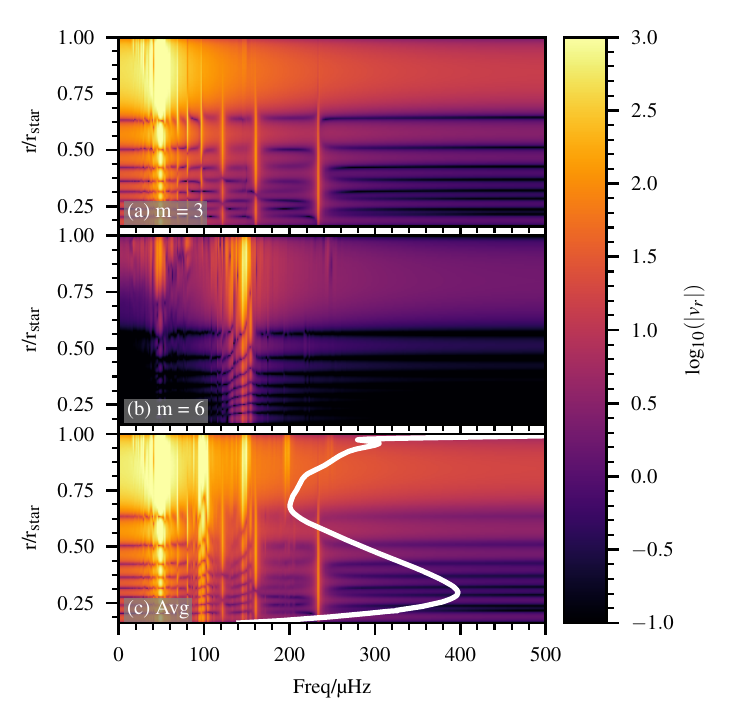}
	\caption{Radial velocities as functions of stellar radius and frequency for a wave forced at 50 \si{\micro}Hz and $m = 3$. Panel (a) shows the radial velocities at a wavenumber $m = 3$, panel (b) is for $m=6$ and panel (c) shows the wavenumber-averaged radial velocities. The white line in panel (c) indicates the \brunt{} frequency profile for this stellar model \label{fig:trfft_single_wave_50}}
	\centering
\end{figure}

Fixing the wavenumber at $m = 3$, we study the evolution of the radiation zone with two other forced frequencies: one at 10 \si{\micro}Hz and the other at 50 \si{\micro}Hz. In both cases, the forced wave amplitude was set to 1000 cm s$^{-1}$.  Figure~\ref{fig:trfft_single_wave_50} shows the frequency spectra for a wave forced at 50 \si{\micro}Hz, similar to Fig.~\ref{fig:trfft_single_wave}. Looking at the frequency spectrum at $m = 3$ (top panel), we see the waves modes that resonate with the cavity being generated. At $m = 6$ (middle panel), we see a wave at 100 \si{\micro}Hz but no resonant cavity modes are generated. In the bottom panel, we observe peaks at integer multiples of 50 \si{\micro}Hz. This is a clear feature of triadic interaction. Taking two slices of the plot in Fig.~\ref{fig:trfft_single_wave_50} shows that at 90\% the total radius of the star, the frequency slope is -1.931, whilst at 99\%, the slope is -2.072. 
\begin{figure}
	\centering
	\includegraphics[trim={0.0cm 0.0cm 0 0.0cm},clip,width=\columnwidth]{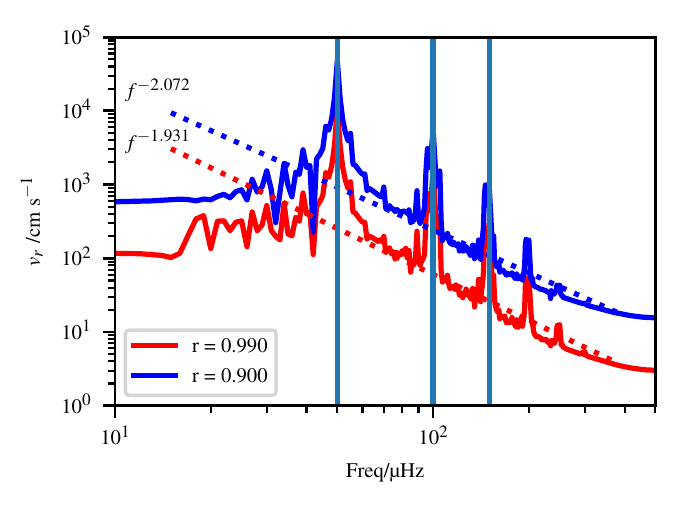}
	\caption{ Temporal Fourier transforms of the radial velocities at three different radii for a wave forced at 50 \si{\micro}Hz and $m = 3$. The radii given in the legend are in units of the total stellar radius. The dashed lines represent straight line fits to $\log_{10} (v_r)$  and $\log_{10}(\mathrm{freq})$. The vertical straight lines represents frequencies 50 \si{\micro}Hz, 100 \si{\micro}Hz and 150 \si{\micro}Hz. \label{fig:single_line_plot_surface_50}}
	\centering

\end{figure}

\begin{figure}
	\centering
	\includegraphics[trim={0.0cm 0.0cm 0 0.0cm},clip,width=\columnwidth]{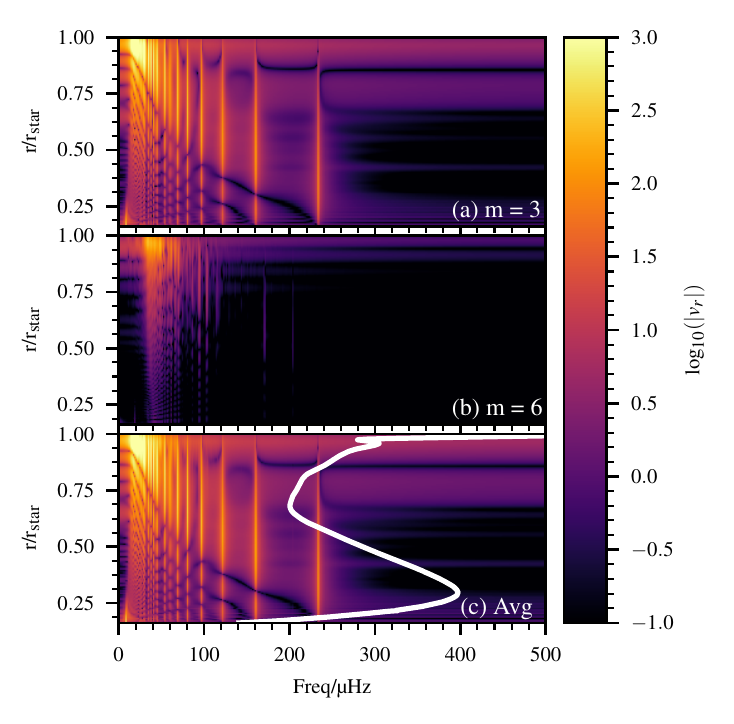}
	\caption{ Radial velocities as functions of stellar radius and frequency for a wave forced at 10 \si{\micro}Hz and $m = 3$. Panel (a) shows the radial velocities at a wavenumber $m = 3$, panel (b) is for $m=6$ and panel (c) shows the wavenumber-averaged radial velocities. The white line in panel (c) indicates the \brunt{} frequency profile for this stellar model.  \label{fig:trfft_single_wave_10}}
	\centering

\end{figure}

\begin{figure}
	\centering
	\includegraphics[trim={0.0cm 0.0cm 0 0.0cm},clip,width=\columnwidth]{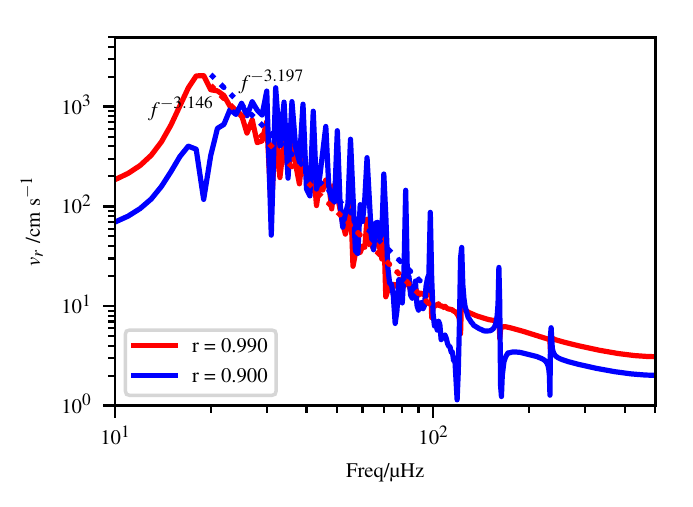}
	\caption{ Temporal Fourier transforms of the radial velocities at three different radii for a wave forced at 10 \si{\micro}Hz and $m = 3$. The radii given in the legend are in units of the total stellar radius. The dashed lines represent straight line fits to $\log_{10} (v_r)$  and $\log_{10}(\mathrm{freq})$. \label{fig:single_line_plot_surface_10}}
	\centering

\end{figure} 

\begin{figure*}
	\centering
	\includegraphics[trim={0.0cm 0.0cm 0 0.0cm},clip,width=\textwidth]{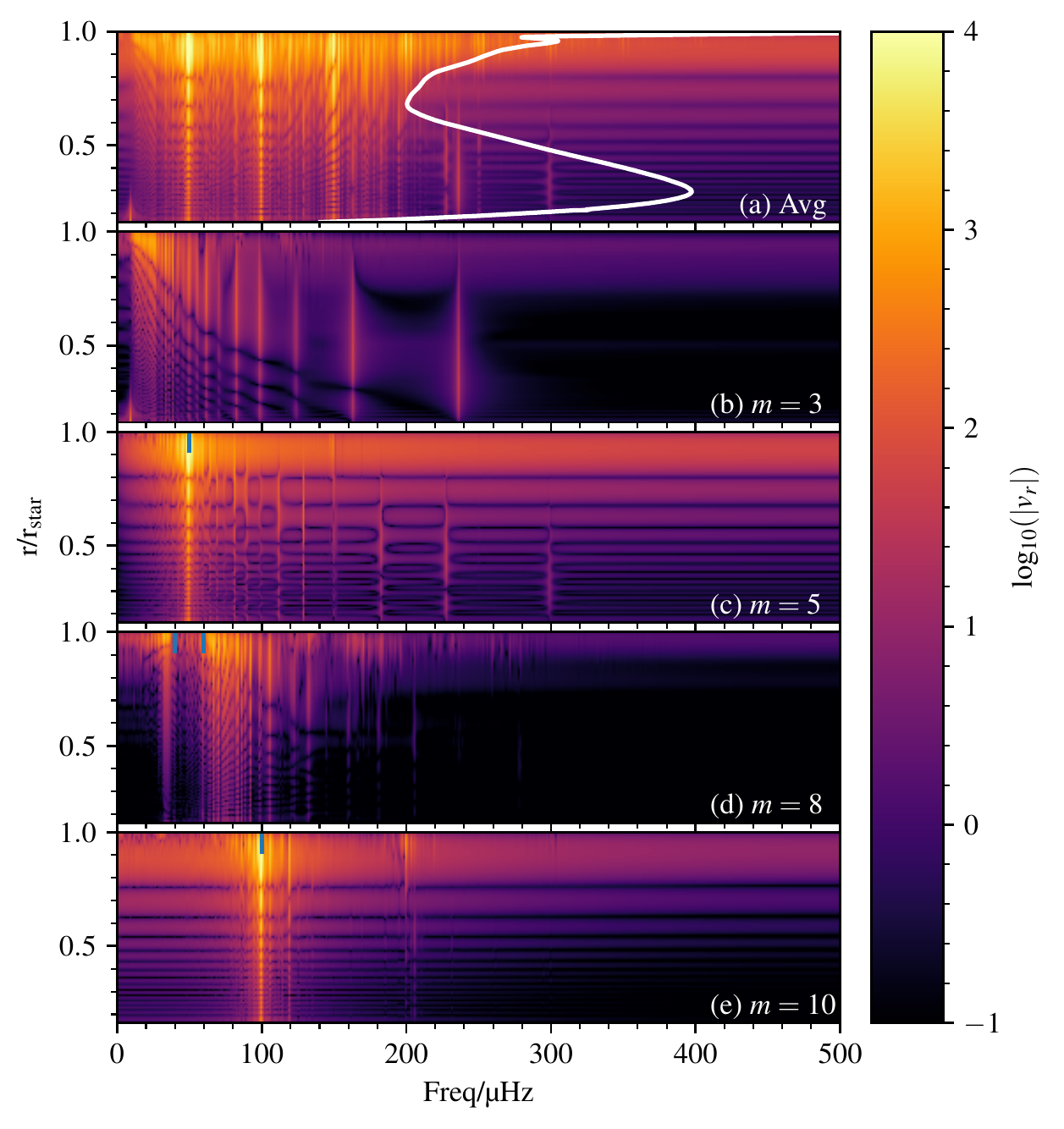}
	\caption{Wave amplitudes as a functions of stellar radius and wave frequency. Panel (a) shows the wave amplitudes averaged over wavelengths with the \brunt{} frequency profile overplotted in white. Panel (b) to panel (e) shows the wave amplitudes at $m=3$, $m=5$, $m=8$, $m=3$ and $m=10$ respectively.  \label{fig:trfft_two_wave_10a50}}
	\centering

\end{figure*}

When a wave with a frequency of 10 \si{\micro}Hz and $m = 3$ is forced at the bottom boundary, this wave can be seen to damp over a very short distance as shown in Fig.~\ref{fig:trfft_single_wave_10}\textcolor{blue}{(a)}. However, at $m = 3$, almost all the cavity modes are forced. At $m = 6$ (panel (b)), triadic interactions close to the bottom boundary allow a wave of 20 \si{\micro}Hz to be formed, which then excites cavity modes at higher frequencies. Also, one can see that at $m = 6$, there is no cavity mode at 100 \si{\micro}Hz, which can be a reason behind the lack of cavity modes in Fig.~\ref{fig:trfft_single_wave_50}\textcolor{blue}{(c)}. Finally, in Fig.~\ref{fig:single_line_plot_surface_10}, we can see that the frequency slopes are -3.197 and -3.146 for the cases of $r = 0.9$ and $r = 0.99$ respectively. This slope is even steeper than that in the case of the 50 \si{\micro}Hz wave.    

To summarise, we found that when single waves are forced at the bottom of the radiation zone with very low nonlinearity parameter, non-linear energy transfer occurs from the forced wave to waves with different frequencies and wavelengths. The two main energy transfer mechanisms are triadic interaction and cavity mode interaction. Moreover, we found that the slope of the frequency spectrum becomes steeper with smaller forced wave frequency. 

\begin{figure}
	\centering
	\includegraphics[trim={0.0cm 0.0cm 0 0.0cm},clip,width=1.0\columnwidth]{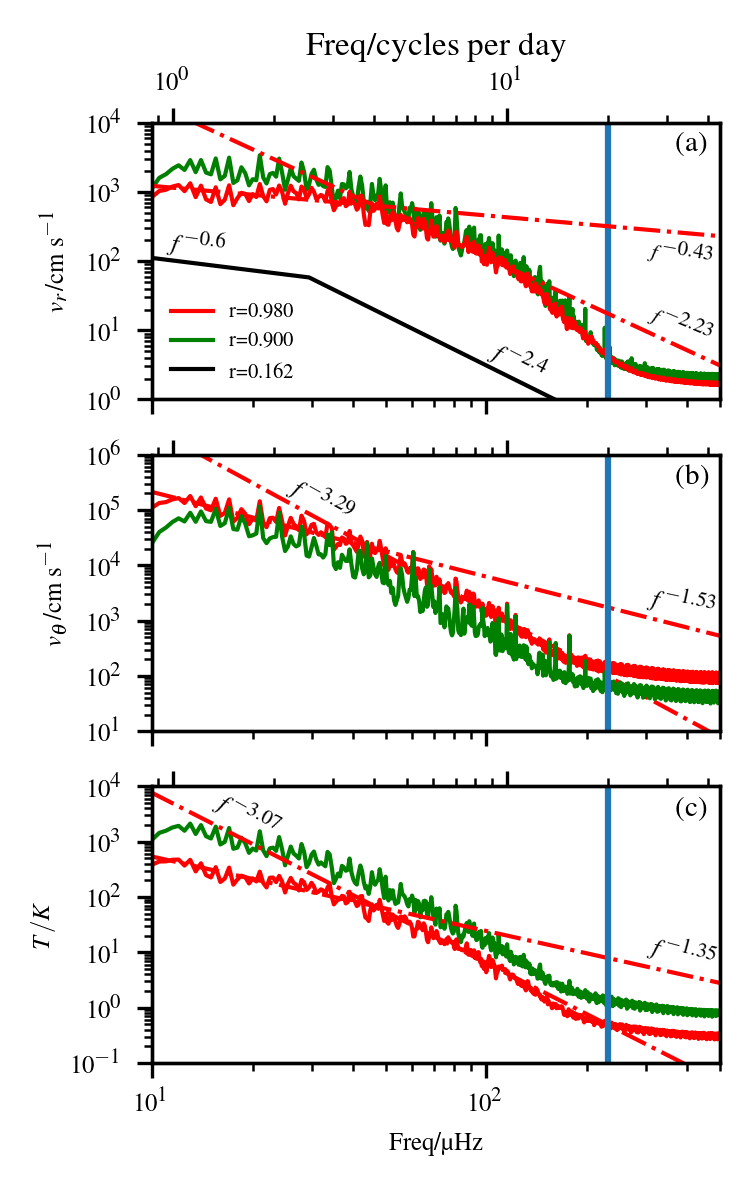}
	\caption{Temporal Fourier transforms of the (a) radial velocities, (b) tangential velocities and (c) temperature perturbations at the generation radius (black line), r = 0.90 (green line) and r = 0.98 (red line), for a 3 M$_{\odot}$ ZAMS star with no rotation. The red dotted-dash lines represent the fit done to the spectrum at r = 0.98 between 10 \si{\micro}Hz and 50 \si{\micro}Hz, and between 50 \si{\micro}Hz and 150 \si{\micro}Hz. The vertical blue lines represent the upper frequency limit in the observational data used in \citet{2015ApJ...806L..33A} to compare against numerical results.  \label{fig:fullspectra_for_3Ms_ZAMS}}
	\centering
\end{figure}

\begin{figure*}
	\centering
	\includegraphics[trim={0.0cm 0.0cm 0 0.0cm},clip,width=1.0\textwidth]{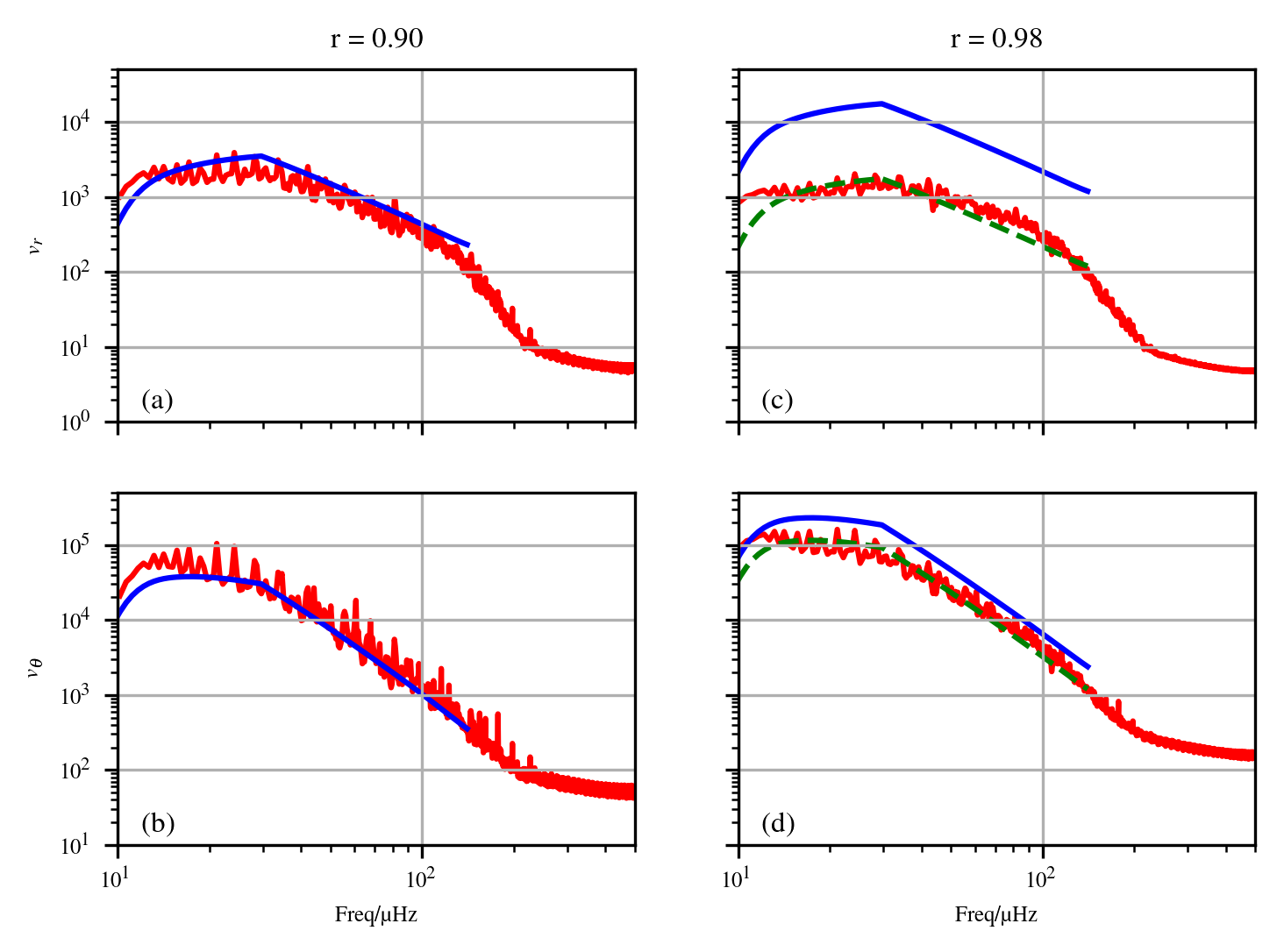}
	\caption{ The radial and tangential velocity spectra at r = 0.9 (first column) and r = 0.98 (second column). The blue lines represent the spectra from linear theory and the green dashed lines represent the linear spectra multiplied by a constant factor to match the non-linear amplitudes.  \label{fig:compare_spectra_for_3Ms_ZAMS}}
	\centering
\end{figure*}

\begin{figure*}
	\centering
	\includegraphics[trim={0.0cm 0.0cm 0 0.0cm},clip,width=0.9\textwidth]{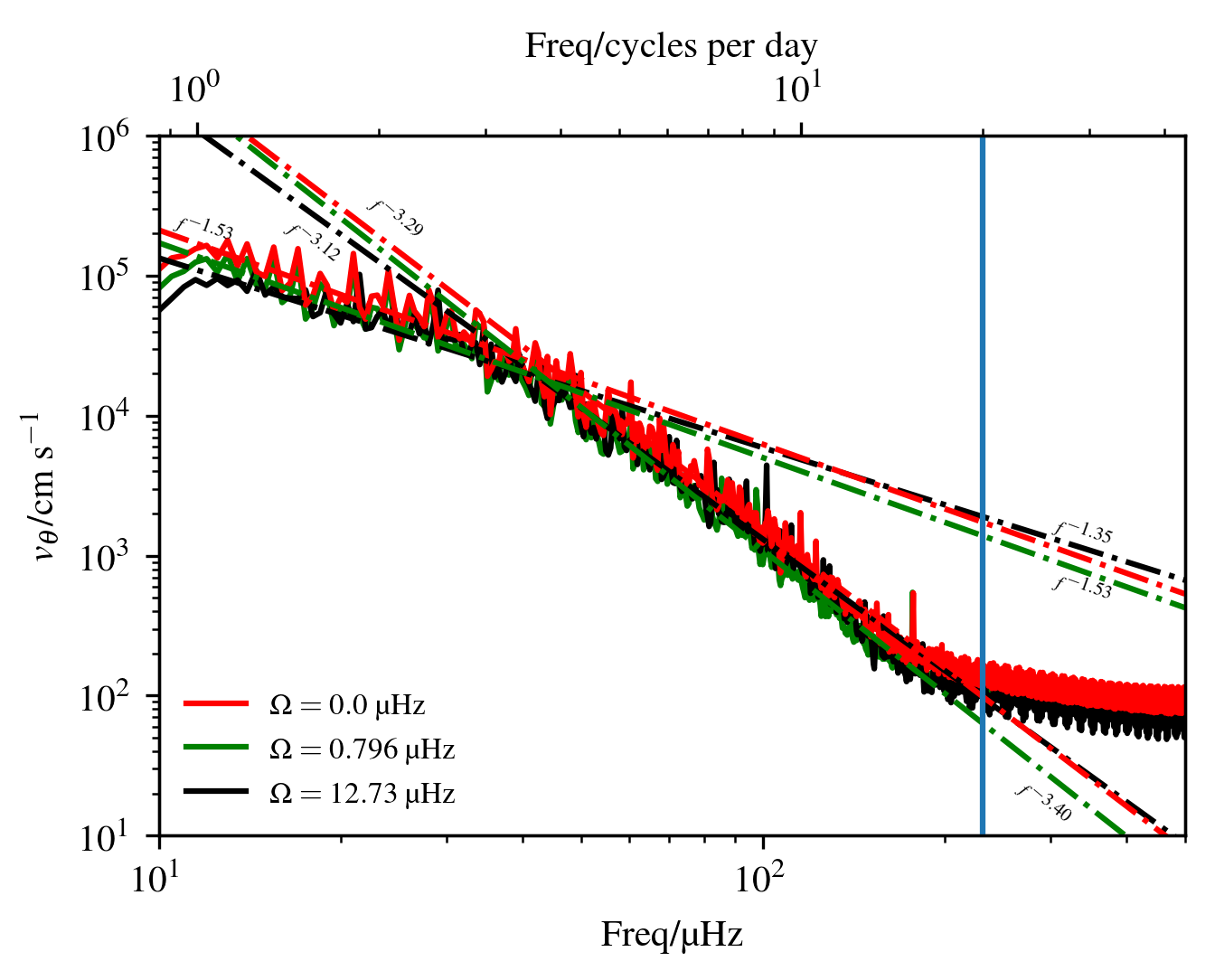}
	\caption{Temporal Fourier transforms of the tangential velocities close to the stellar surface (r = 0.98) for 3 M$_{\odot}$ ZAMS star with no rotation (red line), $\Omega = $ 0.796 $\si{\micro}$Hz (green line) and $\Omega = $ 12.73 $\si{\micro}$Hz (black line). The generation spectrum when $\Omega = $ 0.796 $\si{\micro}$Hz has a slope of -0.5/-2.35 and when $\Omega = $ 12.73 $\si{\micro}$Hz, the slope is -0.25/-1.95. For zero rotation, the slope is the same as $\mathrm{R_{break}}$. The vertical blue line represents the upper frequency limit in the observational data used in \citet{2015ApJ...806L..33A} to compare against numerical results. \label{fig:rot_comparison}}
	\centering
\end{figure*}

\begin{figure*}
	\centering
	\includegraphics[trim={0.0cm 0.0cm 0 0.0cm},clip,width=0.9\textwidth]{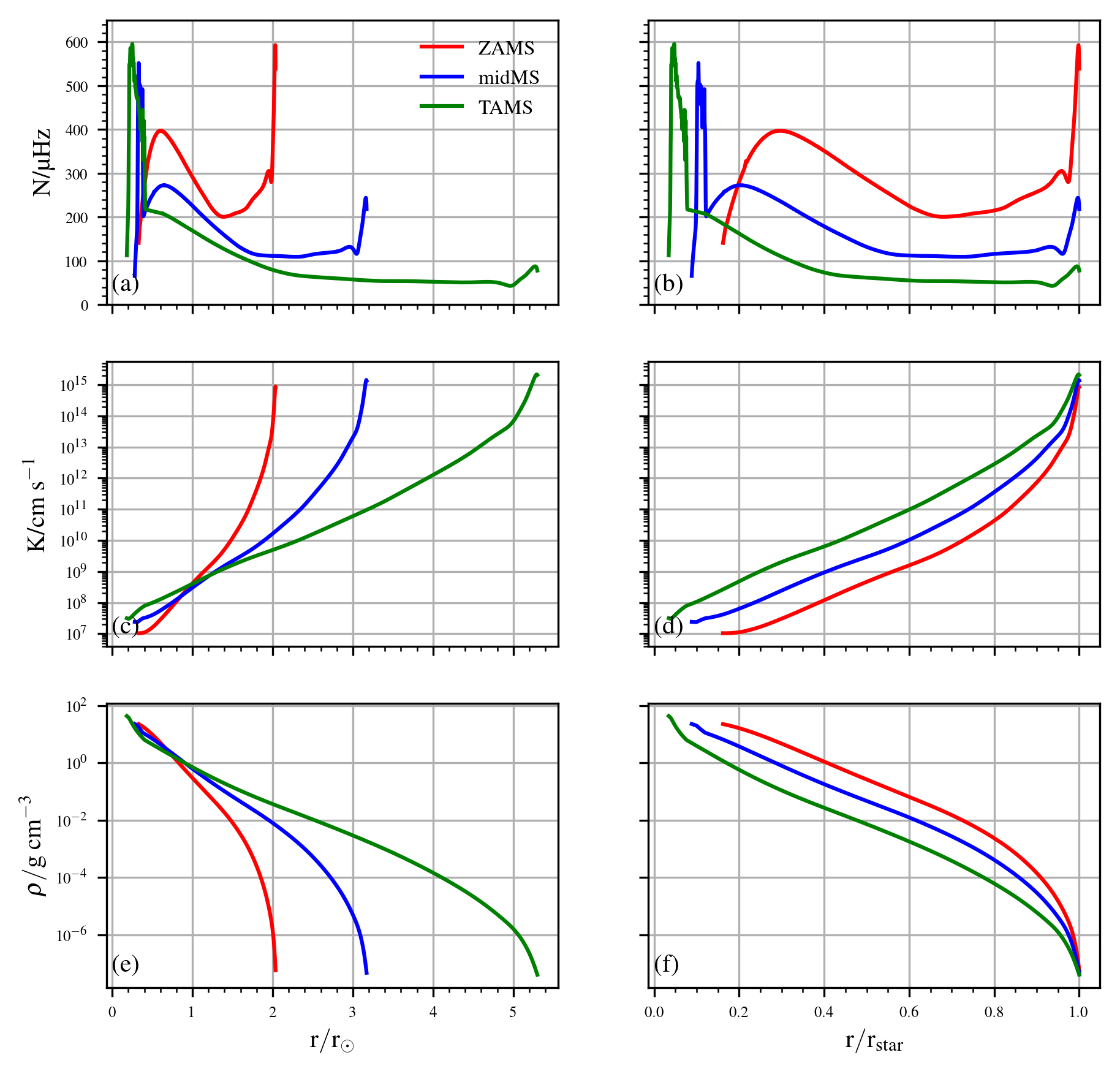}
	\caption{ The \brunt{} frequency (top panel), thermal diffusivity (middle panel) and density (bottom panel) profiles as a function of stellar radius as a unit of total solar radius for stellar models of a 3 M$_{\odot}$ at ZAMS (red line), midMS (blue line) and TAMS (green line).  \label{fig:age_comparison_parameters}}
	\centering
\end{figure*}

\begin{figure*}
	\centering
	\includegraphics[trim={0.0cm 0.0cm 0 0.0cm},clip,width=0.9\textwidth]{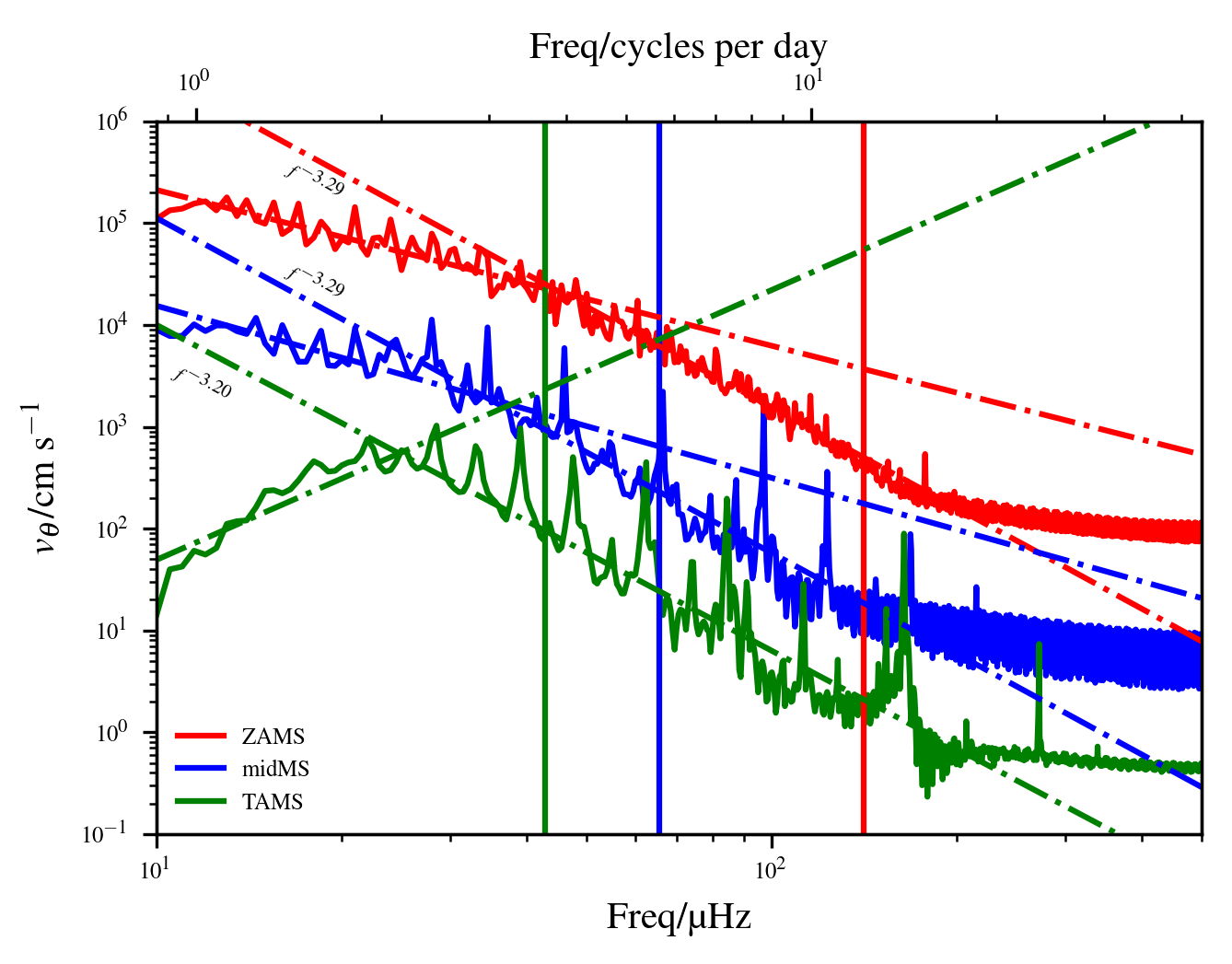}
	\caption{ Temporal Fourier transforms of the tangential velocities close to the stellar surface (r = 0.98) for 3 M$_{\odot}$ ZAMS star (red line), 3 M$_{\odot}$ midMS star (blue line) and 3 M$_{\odot}$ TAMS star (green line). The solid vertical lines represent the minimum \brunt{} frequency of the respective models. The linear fits (dashed-dotted lines) show the general trend of the spectrum and were not done using a least-square fit method.  \label{fig:age_comparison}}
	\centering
\end{figure*}

\begin{figure}
	\centering
	\includegraphics[trim={0.0cm 0.0cm 0 0.0cm},clip,width=1.0\columnwidth]{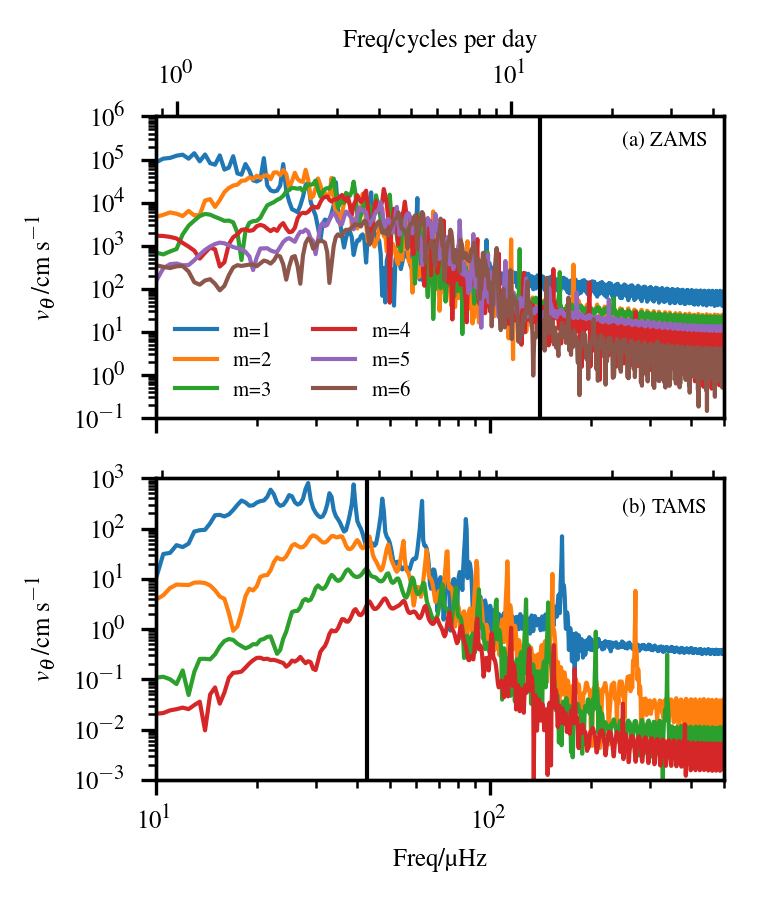}
	\caption{ Tangential velocity amplitudes at different frequencies for horizontal wavenumbers up to 6. Panel (a) represents a ZAMS model whilst panel (b) represents a TAMS model. The vertical black lines represent the minimum \brunt{} frequencies in the respective model.  \label{fig:individual_m_tams_zams}}
	\centering
\end{figure}

\begin{figure}
	\centering
	\includegraphics[trim={0.0cm 0.0cm 0 0.0cm},clip,width=0.5\textwidth]{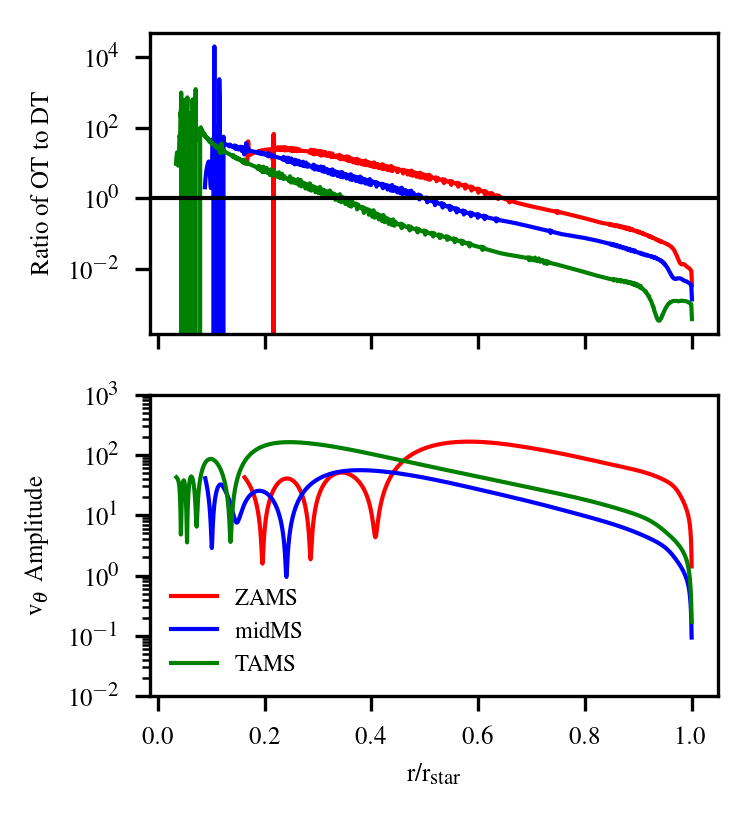}
	\caption{The ratio of the oscillatory term (OT) to the density term (DT) in Eq.~\eqref{eq:linear_vr} (top panel) and the $v_{\theta}$ amplitude as a function of radius in units of total stellar radius for a 40 \micromu Hz, m = 1 wave. The colors represent different stellar models as shown in the legend. The horizontal black line in the top panel represents a ratio of 1. \label{fig:ratio}}
\end{figure}

\subsection{Two-Wave Analysis}
In a system with multiple forced waves, there will be interactions between different waves in addition to wave self-interactions, as discussed in the previous section. These non-linear interactions can lead to generation of waves at different frequencies and wavelengths. We investigate this non-linear wave-wave interaction by forcing two different waves at the bottom of the radiation zone; one at 10 \si{\micro}Hz, $m = 3$ and another at \SI{50}{\micro Hz}, $m=5$. Both waves were forced with the same amplitude of \SI{1000}{cm.s^{-1}}.  

Figure~\ref{fig:trfft_two_wave_10a50} shows the frequency spectra of the wave amplitudes in the radiation zone for different wavenumbers. Looking at the top panel of the figure which shows the wavenumber-averaged wave amplitudes, we observe several distinct features. First, we see clear standing modes at frequencies which are integer multiples of 50 \si{\micro}Hz up to the \brunt{} frequency limit in the domain. These are waves generated from the self-interaction of the forced 50 \si{\micro}Hz wave, as discussed in the the previous section. Figure~\ref{fig:trfft_two_wave_10a50}\textcolor{blue}{(c)} shows that the 50 \si{\micro}Hz forced wave is generated at the correct frequency and additionally, grows in amplitude as it propagates through the radiation zone. 

The second feature of panel (a) is we see the 10 \si{\micro}Hz forced wave being damped over a very short radial distance. Looking at the frequency spectrum at $m=3$ (Fig.~\ref{fig:trfft_two_wave_10a50}\textcolor{blue}{(b)}), we see that the damping of 10 \si{\micro}Hz occurs rapidly in the radiation zone and the pattern of IGWs is similar to that seen in Fig.~\ref{fig:trfft_single_wave_10} for the single forced wave.

Finally, the interaction between the two forced waves produces waves at wavenumbers that are non-integer multiples of 3 and 5. Looking at Fig.~\ref{fig:trfft_two_wave_10a50}\textcolor{blue}{(d)}, which shows the frequency spectrum at $m = 8$, we see waves near 40 \si{\micro}Hz and 60 \si{\micro}Hz spanning the whole simulation domain, which are expected from triadic interaction. We also observe waves at frequencies lower than 40 \si{\micro}Hz and higher than 60 \si{\micro}Hz. One possible explanation for the formation of these waves is cavity mode interaction, as discussed in the previous section. Another explanation is non-linear interaction between secondary waves. Waves formed from non-linear interactions between the forced waves can then interact to form new waves with a different wavenumber and frequency. Thus, we found that non-linear interaction between two different waves leads to the formation of waves at a whole range of other wavenumbers and frequencies.  

\subsection{Multiple-Wave Analysis}
Currently, we do not have a complete understanding on how core convection processes generate a spectrum of gravity waves at the convective-radiative boundary. The spectra described in Section~\ref{sec:gen_spec} are a selection of predictions from different numerical simulations and theoretical studies. To study the effect of these different excitation spectra on the waves in the radiative zone, we forced a range of waves at the bottom boundary of our system with frequencies between 10 \si{\micro}Hz and 500 \si{\micro}Hz and $m = 1 - 20$ \footnote{We only consider waves with frequencies more than 10 \micromu Hz because to resolve waves down to 1 \si{\micro}Hz with four grid points, we need at least 12000 grid points in the radial direction, which is approximately 10 times the current resolution of our simulations.}. 

We found that for the cases of spectrum K and spectrum LD, the slope of the frequency spectrum was so steep that more than 50\% of the energy in waves was in frequencies less than 10 \si{\micro}Hz. Waves with such low frequencies are damped over very short distances, which means that they do not directly contribute to the surface spectrum. In addition to this, spectrum K and spectrum LD were derived for stars with convective envelopes, which makes the work done in \cite{2013ApJ...772...21R} more relevant for this investigation. Thus, we work here with just one generation spectrum; $\mathrm{R_{break}}$.

Figure~\ref{fig:fullspectra_for_3Ms_ZAMS} shows frequency spectra at two different radii for the generation spectrum, $\mathrm{R_{break}}$, where panel (a) shows the radial velocity amplitudes ($v_r$), panel (b) shows the tangential velocity amplitudes ($v_{\theta}$) and panel (c) shows the temperature perturbation amplitudes ($T$), at r = 0.98 (red line) and r = 0.90 (green line). Simulations was run for at least 100 wave cycles, for the wave with the smallest frequency and wavenumber (10 \micromu{}Hz, $m=1$) for all the models in this work and we found that for this wave to reach an amplitude that is constant up to 5\%, the total time elapsed less than 40 wave cycles. This particular wave is chosen as an example here as it takes the longest time to reach equilibrium. We will be using r to represent r/r$_{\mathrm{star}}$ from here on. As mentioned in Section \ref{sec:intro}, previous work on anelastic simulations of stellar interiors do not extend the simulation domain beyond 90\% of the total stellar radius for numerical stability. We have included the spectrum at this radius to be compared with the spectrum at the surface (r = 0.98). For all the quantities shown in the figure, the frequency slopes at r = 0.98 and r = 0.90 are similar. These frequency slopes are negative close to the stellar surface, with the values being $-0.43/-2.23$ (low frequency/high frequency regime)\footnote{The high frequency regime is defined from 50 \si{\micro}Hz to 150 \si{\micro}Hz and the low frequency regime is defined from 10 \si{\micro}Hz to 50 \si{\micro}Hz.} for radial velocities, $-1.35/-3.07$ for temperature perturbations and $-1.53/-3.29$ for tangential velocities. Linear theory shows that the frequency slope for the radial velocities is related to the frequency slope of the tangential velocities by approximately a factor of $\omega^{-1}$ through the following relation:
\begin{align}\label{eq:linear_relationship_vr_vtheta}
	\frac{v_{\theta}}{v_{r}}&=\left(\frac{N^2}{\omega^2}-1\right)^{\frac{1}{2}}\\
						  &\approx \frac{N}{\omega},
\end{align}
where the approximation in the second line is true for frequencies much smaller than the \brunt{} frequency. From the ratio of the exponents we can see that this relation is approximately satisfied in both the high and low frequency regimes in our simulation results. Furthermore, comparing the slope of IGWs seen here with observational results from \citet{dom_igw_exp_range}, we see that our tangential velocity slopes are within the observational prediction range. We have included the upper frequency limit of observational data from \citet{2015ApJ...806L..33A} as a reference point in our plots and we can see that our frequency slopes are for IGW frequencies below this observational limit. This limit has been increased to a larger frequency in \citet{dom_igw_exp_range}.  

Contrary to the expected effect due to the density stratification, both the radial velocity amplitudes in panel (a) and the temperature perturbation amplitudes in panel (c) at r = 0.98 are lower than those at r = 0.90. This is likely an artefact of the impermeable upper boundary condition for these quantities. In panel (b), we see the tangential velocities to be higher at r = 0.98 compared to those at r = 0.90, exhibiting a trend that is opposite to those shown by the radial velocities and temperature perturbations. This is due to the top boundary condition for tangential velocities being stress-free.

To obtain a better understanding of the trend exhibited by the radial and tangential velocities, we compare our results with the work done in \cite{rathish2019}, where a simple linear model was used to investigate the change in IGW amplitude throughout propagation. This model was created by linearising the Navier-Stokes equations and applying the WKB approximation in a spherical geometry with a locally Boussinesq, but globally anelastic approximation, which allows for the effect of damping to be included. In this work, we followed the same procedure in two-dimensional cylindrical coordinates, resulting in Eq.~\eqref{eq:linear_vr}, and applied the WKB approximation to the whole equation. Wave damping is assumed to be caused by viscosity or radiative diffusion, depending on which effect is stronger. Figure~\ref{fig:compare_spectra_for_3Ms_ZAMS} shows the radial and tangential velocity spectra at two different radii from the non-linear simulation (red line) and linear model (blue line).

At r = 0.90, the non-linear spectra for $v_r$ and $v_{\theta}$ matches the linear spectra very well, in terms of the expected frequency slope, as seen in panel (a) and (b). At  r = 0.98, there is a clear mismatch between the linear (blue lines) and non-linear (red lines) amplitudes for both the radial and tangential velocities. As discussed above, the IGW amplitudes near the surface are reduced due to the boundary conditions. However, we see that the slopes match remarkably well, from the green lines, which show the linear model predictions multiplied by a certain factor so that it overlays the numerical simulation results. The linear model has a cut-off at 140 \si{\micro}Hz, which is the smallest \brunt{} frequency in the simulation domain. At the lowest frequencies, there is significantly more energy in the full non-linear simulation spectrum, which is likely due to local non-linear energy transfer (as seen in Fig.~\ref{fig:trfft_single_wave}\textcolor{blue}{(a)} and Fig.~\ref{fig:single_line_plot_surface}).  

\subsubsection{Rotation and Age Analysis}

Convective processes are known to be affected by rotation. In the case of stellar convection, \cite{2013ApJ...772...21R} showed that the spectrum of gravity waves due to core convection for a 3 $M_\sm{}$ varies when rotation is introduced to the system. We ran non-linear simulations with two of the rotational velocities and their corresponding wave spectra from that paper to study the effect of rotation on gravity waves in the radiation zone. Note that we consider the tangential velocity to be the best proxy to compare with observational results (which mainly work with surface brightness variation) and not temperature, as the relation between temperature variation and surface brightness variation is not well-understood. Thus, we only show tangential velocity results from here on.

The stellar angular velocities used here are 0.796 $\si{\micro}$Hz and 12.73 $\si{\micro}$Hz, referred to as models U3 and U8 in \citet{2013ApJ...772...21R}. The critical stellar angular velocity is 60 $\si{\micro}$Hz, so our fastest rotating model is at 20\% of the critical velocity or break-up velocity. For $\Omega = $ 0.796 $\si{\micro}$Hz, the generation spectrum has a frequency slope of $-1.43/-2.27$ and for $\Omega= $12.73 \micromu Hz, the generation spectrum has a frequency slope of $-1.07/-2.36$. For zero rotation, the slope is $-1.55/-2.41$. Figure~\ref{fig:rot_comparison} shows the tangential velocity spectra at $r = 0.98$ for three different stellar angular velocities, which are $\Omega = $ 0.0 $\si{\micro}$Hz (red line), $\Omega = $ 0.796 $\si{\micro}$Hz (green line) and $\Omega = $ 12.73 $\si{\micro}$Hz (black line). In all cases, the surface spectra for the tangential velocities were found to have a steeper slope than the those at generation. However, comparing the slopes at different stellar angular velocities, we found all slopes at both the low and high frequency ranges to be very similar. 

To investigate IGWs in stellar models at different ages, we begin by noting that the work done in \cite{rathish2019} showed that at different stellar ages, the slope of the frequency spectrum at the surface is dependent mainly on the generation spectrum. Here, we have kept the generation spectrum constant for all three models of a 3 $M_\sm{}$: Zero-Age Main Sequence (ZAMS), middle-of-the-Main-Sequence (midMS) and Terminal-Age-Main-Sequence (TAMS). This fixes the radial velocity spectrum at the inner boundary for all three models but not the tangential velocity spectrum. Locally, tangential velocities and radial velocities are related by Eq.~\ref{eq:linear_relationship_vr_vtheta}. Thus, the tangential velocity generation spectrum varies between the different models by the same amount the \brunt{} frequency varies between the models at generation. However, in Fig.~\ref{fig:age_comparison_parameters}, which shows the \brunt{} frequency ($N$), thermal diffusivity ($K$) and density ($\rho$) profiles of all three stellar models as functions radius in units of the solar radius (r$_{\odot}$) and in units of total stellar radius (r$_{\mathrm{star}}$), panels (a) and (b) show that there is little difference between the \brunt{} frequencies at generation for the different stellar models, leading to the tangential velocity profiles at generation being similar too.  

Figure~\ref{fig:age_comparison} shows the surface tangential velocity spectra, defined at 98\% of the total stellar radius, for the three different stellar ages. There are a few distinct features seen in this plot. The first feature is that the amplitudes of all waves at the surface decrease as the star ages. This occurs due to a few different factors. The first factor is that older stars are larger causing IGWs to travel farther from generation and reach the surface with lower amplitudes than those in younger stars. This spreading out of waves is expected to cause the IGW amplitudes to decrease as $r^{-1}$ in 2D. The second factor is that all IGWs experience extra damping due to them losing their wave-like properties at the turning point (see Section~\ref{sec:linear_theory}) in the radiation zone. For older stars, the turning point is located at a lower fraction of their total radius. To understand this better, we look at the top panel of Fig.~\ref{fig:ratio}, which shows the ratio between the oscillatory terms(OT) and the remaining density terms (DT) in Eq.~\eqref{eq:linear_vr} as functions of the total stellar radius for a 40 \micromu Hz, m = 1 wave in different stellar models. The turning point location is indicated by the ratio = 1 (horizontal black line). We can observe that the wave maintains its oscillatory property (indicated by the ratio > 1) for a larger proportion of the radiation zone in younger stars. This can also be observed in the bottom panel of Fig.~\ref{fig:ratio}, which shows the horizontal velocity amplitude at 40 \micromu Hz, m = 1, for different stellar models, where the waves initially show an oscillatory behaviour followed by a decaying behaviour. 

The second feature of Fig.~\ref{fig:age_comparison} is the stationary mode peaks become more prominent for older stars. This effect is caused by the decreasing average \brunt{} frequency as the star ages, as seen in panels (a) and panels (b) of Fig.~\ref{fig:age_comparison_parameters}. As a result, the high wavenumber waves undergo rapid damping causing only the low wavenumber waves to reach the surface. This can be seen in Fig.~\ref{fig:individual_m_tams_zams}, which shows the surface IGW spectra at different horizontal wavenumbers, for the (a) ZAMS model and the (b) TAMS model. Waves with horizontal wavenumbers up to $m = 6$ are still contributing to the total spectrum for a ZAMS star (panel (a)) whilst only waves up to $m = 3$ contribute to the total spectrum for TAMS star (panel (b)). In fact, this is also a contributing factor to the decreasing amplitudes of all waves in older stars, as discussed in the previous paragraph.    

A third distinguishing feature appears at lower frequencies (12 \micromu Hz $\leqslant$ $f$ $\leqslant$ 20 \micromu Hz). The surface IGW spectrum for the TAMS model shows a positive slope whilst those for the ZAMS and midMS models show negative slopes. This is because low-frequency waves are damped more efficiently by the broad \brunt{} frequency spike near the interface, as seen in Fig.~\ref{fig:age_comparison_parameters}\textcolor{blue}{(a)} for the TAMS model. 

The final feature of Fig.~\ref{fig:age_comparison} is that the frequency slopes at high frequencies (50 \micromu Hz to 150 \micromu Hz) are similar for all models close to the stellar surface. They were found to be one order of magnitude larger than the frequency slope for the radial velocities. The surface radial velocity slopes were found to be similar those at generation.  Note that although the linear fits for these spectra were done visually as least-square fits are perturbed by the strong peaks, observational methods such as pre-whitening involves the analysis of stellar IGW spectra without these peaks. Thus, we find that our result is consistent with observational results \citep{2015ApJ...806L..33A,2018MNRAS.480..972R,dom_igw_exp_range}. For the list of all the frequency slopes for the stellar models in this paper, please refer to Table~\ref{table:slopes}.

\begin{table}
	\centering		
	\begin{tabular}{ccc}
		\hline \hline
		Frequency Slopes & Low-Frequency & High-Frequency \\ \hline \hline 
		ZAMS ($\Omega$ = 0.0 \si{\micro}Hz) & & \\ 
		Generation Spectra & -1.55 & -2.4 \\
		Surface Spectra & -1.53 & -3.29 \\ \hline
		
		ZAMS ($\Omega$ = 0.796 \si{\micro}Hz) & & \\ 
		Generation Spectra & -1.43 & -2.27 \\
		Surface Spectra & -1.53 & -3.40 \\ \hline
		
		ZAMS ($\Omega$ = 12.73 \si{\micro}Hz) & & \\ 
		Generation Spectra & -1.07 & -2.41 \\
		Surface Spectra & -1.35 & -3.12 \\ \hline
		
		midMS   &  &  \\ 
		Generation Spectra & -1.55 & -2.4 \\
		Surface Spectra & -1.69 & -3.29 \\ \hline
		
		TAMS    & & \\ 
		Generation Spectra & -1.55 & -2.4 \\
		Surface Spectra & 5.10 & -3.20 \\ \hline
		
	\end{tabular}
	\caption{The table shows the frequency slopes of the tangential velocity amplitudes for all the models used in this work. The generation spectra is defined at the bottom boundary of the radiation zone and the surface spectra is defined at r = 0.98 $r_{\mathrm{star}}$.
	\label{table:slopes}}
\end{table}

\section{Conclusion and Discussion}
We ran two-dimensional simulations of the evolution of stellar radiation zones, in which IGWs were forced at the bottom boundary. Since the convection zone was not included, we are able to simulate the radiation zone up to 99\% of the stellar radius and implement realistic thermal diffusivities. We started the investigation with a series of validation tests by investigating monochromatic IGW forcing. We found that for different forced wave frequencies, the waves self-interact through triadic interactions and also resonate with stationary modes within the cavity if the forced wave frequency is close to a stationary mode frequency. 

Our main conclusion is we find that the surface IGW frequency slopes broadly match those predicted by linear theory and reflect the generation spectrum. Furthermore, we find that up to 20\% the critical velocity of a star, stellar rotation (in this case, solid body rotation) does not affect the surface IGW frequency slope. For these simulations, we took into account the effect of rotation on both the slope of the generation spectrum, using results from \cite{2013ApJ...772...21R} and the rotation of the radiation zone.

Another result of our investigation is that the surface IGW frequency slopes are similar at different stellar age. Note that we used the same generation spectrum for three different stellar models (i.e. ZAMS, midMS and TAMS). We also find that for older stars, the IGW amplitudes are not only lower, they have more distinct peaks in their spectra.  

In summary, our results show that the surface frequency slopes are consistent with the 2D hydrodynamical simulations of stellar interiors \citep{2013ApJ...772...21R} and observational results to-date \citep{2015ApJ...806L..33A,2018MNRAS.480..972R,dom_igw_exp_range}. We find the most recent observational work, \cite{dom_igw_exp_range}, which found that the frequency slope magnitudes of IGW signatures from a variety of blue supergiants (BSGs) are less than 3.5, to be the most compelling comparison to our results. This is due to the large number and variety of stars (i.e. O-- and B-- type stars), and their argument that the variability of the surface IGW spectrum does not have any dependence on metallicity.    

However, there are a few differences between our models and blue supergiants, which requires further investigation. First, BSGs are more evolved and more massive. Both stellar age and mass has been shown to cause high variability in observed surface IGW spectra and we expect the same to happen at generation. Thus, studying the effect of age and mass on generation spectra will be highly beneficial. Second, the chemical composition, size and rotational speed of these BSGs are expected to be different from our models. With chemical composition and size, we expect IGWs to be damped more in these supergiants but the frequency slope should not affected. However, BSGs possess differential rotation, which is expected to cause frequency shifts in the surface IGW spectrum. Finally, BSGs are three-dimensional structures, although 3D simulations do not promise more accurate results in the framework of this paper (i.e. not including the convection zone), extension of this work to 3D can allow us obtain a better understanding of wave propagation in 3D. Mainly, as shown in Section~\ref{sec:linear_theory}, linear theory predicts that the surface IGW amplitudes for 3D should decrease by a factor of $\omega^{-0.5}$ compared to 2D. 

\section*{Acknowledgements}
We acknowledge support from STFC grant ST/L005549/1 and NASA grant NNX17AB92G. Computing was carried out on Pleiades at NASA Ames, Rocket High Performance Computing service at Newcastle University and the DiRAC@Durham facility managed by the Institute for Computational Cosmology on behalf of the STFC DiRAC HPC Facility (www.dirac.ac.uk), which is funded by BEIS capital funding via STFC capital grants ST/P002293/1 and ST/R002371/1, Durham University and STFC operations grant ST/R000832/1. We thank Conny Aerts, Daniel Lecoanet, May Gade Pedersen and Dominic Bowman for useful conversations leading to the development of this manuscript.

\section*{Data Availability}
Data available on request.


\bibliographystyle{mnras}
\bibliography{bi} 




\bsp	
\label{lastpage}
\end{document}